\documentclass[aps,elsarticle,onecolumn,eqsecnum,nofootinbib]{revtex4}

\usepackage{epsfig}
\usepackage{graphicx}
\usepackage{dcolumn}
\usepackage{amsmath}
\usepackage{enumerate}

\def\gp{$(\gamma_{1}, \gamma_{2})$}
\def\td{$(\theta_{13},\delta)$}
\def\s2{$\sin^{2}2\theta_{13}$}
\def\t13{$\theta_{13}$}
\def\d{$\delta$}

\begin{document}

\begin{flushright}
 IPPP/09/90 \\
DCPT /09/180\\
EURO$\nu$-WP6-09-13    
\end{flushright}

\title{CP-violation reach of an electron capture neutrino beam}

\author{Christopher Orme}

\affiliation{Institute for Particle Physics Phenomenology, Department of Physics, Durham University,
  Durham DH1 3LE, United Kingdom}

\begin{abstract}
   This article extends the work of Bernabeu and Espinoza by examining the CP-violation reach
of a $^{150}$Dy electron capture beam through the variation of the two Lorentz boosts, the number of
useful electron capture decays, the relative run time of each boost and the number of atmospheric
backgrounds. The neutrinos are assumed to be sourced at CERN with an upgraded SPS 
and are directed towards a 440 kton Water Cerenkov detector located
at the Canfranc laboratory. Two large `CP-coverage' choices for the boost pairings are found;
a $\delta$-symmetrical coverage for $(\gamma_{1}, \gamma_{2})$ = (280, 160) and an $\delta$-asymmetric coverage for $(\gamma_{1}, \gamma_{2})$ =
(440,150). With a nominal useful decay rate of $N_{\rm ions} = 10^{18}$ ions per year, the $\delta$-symmetric setup can
rule out CP-conservation down to $\sin^{2}2\theta_{13} = 3\cdot 10^{-4}$. To reach $\sin^{2}2\theta_{13} = 1\cdot 10^{-3}$ for both $\delta < 0$ and
$\delta > 0$ requires a useful decay rate of $N_{\rm ions} = 6\cdot 10^{17}$ ions per year.

\end{abstract}

\maketitle

\section{Introduction}

  The results of a series of ground breaking neutrino experiments indicate that neutrinos are both massive and that
they mix amongst themselves. Data from atmospheric~\cite{SKatm,atm}, solar~\cite{sol,SKsolar,SNO}, reactor~\cite{CHOOZ,PaloVerde,KamLAND} 
and long-baseline accelerator~\cite{K2K,MINOS} neutrino experiments can be accommodated by two approximate 2-neutrino mixing schemes, each parameterised
by a mass-squared splitting and a mixing angle.

  The goal of the future experimental program in neutrino oscillations is to determine how these `solar' and `atmospheric' sectors combine into an overall 3-neutrino mixing scheme. This requires a search for sub-dominant appearance
events; a signal whose strength is controlled by a third angle, \t13, that is currently only bounded from above. Up
to three physical phases can be incorporated into a 3-neutrino mixing scheme; however, only the effect of the Dirac
phase, \d, can be seen in an oscillation experiment as CP-violation. If the third mixing is zero then the two 2-neutrino
mixing schemes are disjoint and there is no discrepancy between the neutrino and anti-neutrino mixing matrices

   Defining $\Delta m^{2}_{ji}= m^{2}_{j}- m^{2}_{i}$ , the current best fit values for the oscillation parameters are~\cite{STV08}
\begin{alignat}{4}
&\vert \Delta m_{31}^{2}\vert = 2.4 \times 10^{-3}\:\: {\rm eV}^{2} &\quad \mbox{and} \quad & \sin^{2}\theta_{23} =0.50~; &\notag\\
&\Delta m_{21}^{2} = 7.65 \times 10^{-5}\:\: {\rm eV}^{2} &\quad \mbox{and} \quad & \sin^{2}\theta_{12} =0.304~.& \notag 
\end{alignat}
The third mixing angle, $\theta_{13}$, is constrained to be~\cite{STV08}
\begin{equation}
 \sin^{2} \theta_{13} < 0.040\:\:\: (0.056)\quad \mbox{at}\quad 2\sigma\:\:\: (3\sigma)~;
\end{equation}                                      
although some collaborations report a hint at low significance from a combined analysis of atmospheric, solar and
long-baseline reactor neutrino data~\cite{Fogli:2008jx,Ge:2008sj}.

   Running and near future neutrino oscillation experiments~\cite{MINOS,freactors,T2K,newNOvA} will be the first to probe $\theta_{13}$ below the current
limit and possibly confirm the reported hint for $\theta_{13} = 0$. If $\theta_{13}$ is beyond the reach of near future experiments,
intense sources of neutrinos will be necessary: next generation SuperBeams~\cite{SB_next}, Neutrino Factories~\cite{nufact,nufactlow} and Beta
Beams~\cite{zucchelli,mauro}. The aspiration of future long baseline neutrino oscillation experiments is to determine the third
mixing angle, $\theta_{13}$, whether there is CP-violation in the lepton sector; and to determine the sign of $\Delta m^{2}_{31}$. This is not
straightforward; the analysis of the appearance channels is capable of returning up to 8 equally allowable fits to the
data~\cite{deg1,deg2,deg3,deg4,deg5}. The challenge is therefore successful resolution of these degeneracies and a simultaneous push for the
best sensitivity.

   A Beta Beam is a source of intense and collimated beams of electron neutrinos or anti-neutrinos originating from the
decay of boosted radioactive ions. The proposal is an integrated facility combining individual production, acceleration
and storage components. Ion production could make use of the Isotope Separation Online (ISOL) technique~\cite{ions_ISOL} within
the EURISOL at CERN; production rings~\cite{ions_prod} to minimise losses of parent nuclides in a beam dump; or a two target
system using an intense beam of deuterons~\cite{ions_twotar}. The acceleration would make use of existing or potential upgrades
to current facilities such as the CERN accelerator chain, the Fermilab Main Injector and Tevatron; and maybe DESY.
A storage ring whose straight sections point towards the far detectors will be the main infrastructure addition required.

It is advantageous to produce a clean, collimated beam by this technique since a single boost is all that is necessary
for a laboratory neutrino flux to span an energy range up to several GeV. However, one is not free to choose the
laboratory neutrino spectrum; the boost determines the maximum neutrino in the laboratory frame which in turn determines
the un-oscillated event rate at all energies. One way to alleviate this problem is to use a source of mono-energetic neutrinos so that the boost of the ion can
be altered to freely choose the laboratory frame energy of the neutrinos. Clearly, such an approach needs at least two
boosts as both $\theta_{13}$ and $\delta$ are unknown. 

The use of electron capture decays to source mono-energetic neutrino fluxes
has taken on two guises. The first paper~\cite{sato} (and a subsequent study~\cite{satorolinec}) proposed the use of ions with Q-values less than
twice the electron mass, hence positron decay is kinematically forbidden. These are not realistic, however, since the
ion half-lives are too long and the boosts required can only be achieved with the LHC. The idea of using electron capture decays was in fact first suggested
at the EURISOL meeting in January 2005~\cite{BB_meet} with focus on higher Q-valued ions. This was later expanded into studies for the CERN-Frejus and
CERN-Canfranc baselines~\cite{catalina1,catalina2}. For ions with Q-value greater than twice the electron mass, the electron capture
channel competes with a positron decay `background'; the branching ratio for electron capture channels drops sharply
as the Q-value increases. With Q-values $\sim$ 3-4 MeV and a 1 TeV accelerator, one would be able to place a mono-energetic 
neutrino beam on first oscillation maximum for baselines up to 1000 km, making the idea an attractive one
in Europe with the source based at CERN. However, the positron decay background would make the idea unattractive
since the number of useful electron capture decays would not be high enough for a competitive physics reach. It could
be argued that the positron decay neutrinos could be used to boost sensitivity by providing coverage of the second
oscillation maxima with concentration of the electron capture neutrinos on or around first oscillation maximum.
This is the hybrid Beta Beam strategy explored in detail in~\cite{ECBB}. A number of nuclides that decay quickly through large Gamow-Teller
resonances have been discovered in the Gd region of the Segre chart, see for example~\cite{Ybcom}, that opens the possibility of a higher Q-value electron 
capture beam. These ions source dominant electron 
capture channels at higher
Q-values. One of these ions ($^{150}$Dy) forms the basis of this paper as it did~\cite{catalina1, catalina2} where it was used to source the CERN-Frejus and
CERN-Canfranc baselines with a large Water Cerenkov as the far detector.

   In this article, the work of Bernabeu and Espinoza~\cite{catalina2} is extended to include the effect of the choice of Lorentz
boost pairing, the relative run times at each boost, the number of useful ion decays, and the number of atmospheric
neutrino events. In Sec.~\ref{S:concept}, the electron capture beam concept is reviewed with some discussion on the choice of ions
and some of the technological difficulties not present in a standard Beta Beam. In Sec.~\ref{S:ener} a variant on the intrinsic
degeneracy, which I refer to as the energy degeneracy, is introduced which results from the combination of events
at two energies for a neutrino or anti-neutrino run only. The analyses carried out are summarised in Sec.~\ref{S:analysis}; the
variation of the boosts is performed in Sec.~\ref{S:results}; and the number of useful decays and atmospheric backgrounds are
included in Sec.~\ref{S:bck}. To finish, the use of mono-energetic neutrinos from bound beta decay is discussed in Sec.~\ref{S:BBD},
and the conclusions are drawn in Sec.~\ref{S:con}.

\section{The electron capture beam concept}\label{S:concept}

   Electron capture is a decay channel available to proton-rich nuclei and it competes with positron decay depending
on the energy available. For $Q < 2m_{e}$, positron decay is kinematically forbidden and so electron capture decays form
the entire phase space. For electron capture decays, the rate is proportional to the $Q^{2}$ from the two body decay
and $Z^{3}$ from the square of the electron orbital wave-function; where $Q$ is the Q-value of the decay and $Z$ is the proton number.  
The relative rate with respect to positron decay for $Q > 2m_{e}$ is approximately given by
\begin{equation}
\frac{\Gamma_{\rm EC}}{\Gamma_{\beta}} \propto \frac{(\alpha Z)^{3}}{Q^{3}}~,
\end{equation}
where the matrix elements for the channels are assumed to be identical, $\alpha$ is the fine structure constant,
and the leading $Q^{5}$ is taken for $\Gamma_{\beta}$. For
electron capture with Q-value $Q_{\rm EC}$ and boost $\gamma$, the neutrino flux in the laboratory frame at baseline L from source
is given by~\cite{sato,catalina1,catalina2}
\begin{equation}\label{E:flux}
\frac{dN}{d\Omega dE_{\nu}} = \frac{N_{\rm ions}}{\pi L^{2}}\:\gamma^{2}\:\delta(2\gamma Q_{\rm EC}-E_{\nu})\equiv \Phi(E_{\nu})\:\delta(2\gamma Q_{\rm EC}-E_{\nu})~.
\end{equation}
Trivially, the baselines available to electron capture machines are dependent on the Q-value of the ion and the
maximum boost allowed from the acceleration.

\begin{table}
\begin{center}
\scalebox{1.0}{%
\begin{tabular}{c|c|c|c|c|c}
\quad Parent nucleus \quad & \quad Half-life \quad& \quad EC BR \quad & \quad EC intensity \quad & 
                                                      \quad Ex Daughter level (keV) \quad & \quad Q-value (keV)\quad  \\
\hline
\hline
$^{148}$Dy & 3.1 m & 100 \% & 92.5 \% & 620 & 2678 \\
$^{148}$Er & 4.6 s & 100 \% & 8.8 \% & 0.00 & 6800 \\
$^{150}$Er & 19 s & 100 \% & 59.5 \% & 476+X & 4108 \\
$^{150}$Dy & 7.2 m & 64 \% & 64 \% & 397+Y & 1794 \\
$^{152}$Yb & 3.1 s & 100 \% & 29 \% & 482 & 5470 \\
$^{154}$Er & 3.7 m & 99.53 \% & 96.8 \% & 26.9 & 2032 \\
\end{tabular}
}
\end{center}
\caption{Candidate electron capture beam ions. Based on a similar table from~\cite{Ybcom}.}\label{T:ions}
\end{table}
   A collection of ions in the Gd region of the Segre chart, suitable for European baselines, are presented in Tab.~I.
$^{150}$Dy was chosen in~\cite{catalina1,catalina2} as the resonance did not have a width and because of its relatively low Q-value. From
Eq. 2.2, the neutrino flux is proportional to the square of the boost for a fixed baseline and number of useful decays.
Lower Q-values require higher $\gamma$ to achieve the same laboratory neutrino energies, and so lower Q-values are favoured
in this sense, provided the accelerator is capable of achieving the boosts. The discovery of ions far from the stability
line that decay through a giant Gamow-Teller resonance opened up the possibility of using 1 TeV machines such as an
upgraded Super Proton Synchrotron (SPS)~\cite{LHC_upgrade} (envisaged in some LHC-upgrade scenarios), 
to source the intermediate baselines in Europe. As mentioned 
previously,~\cite{catalina1,catalina2} considered this probability for the CERN-Frejus (130 km) and CERN-Canfranc (650 km) baselines with $^{150}$Dy
as the chosen ion. Since only a neutrino flux is available in an electron capture setup, the approach was to exploit
the different energy dependence of the CP-odd and CP-even properties of the appearance probability (Eq. 3.2) by
running at two different boosts. The purpose of those studies was not to identify Lorentz boosts of particular interest, 
rather to demonstrate the phenomenological feasibility of the idea: the use of multiple boosts to extract the
energy dependence of the event spectra. The goal of this article is to examine the CP-violation reach as a function of
the boosts, number of useful decays and atmospheric event rate.
 
   For electron capture and Beta Beam setups, the parent ions will be accelerated in the existing or upgraded CERN
infrastructure before accumulated and stored in a ring whose long, straight sections source the neutrino flux. For
electron capture machines with ion boost $\gamma$, the mono-energetic neutrino flux at a detector distance L from the source
is given by Eq.~\ref{E:flux}. Since the neutrino energy in the laboratory frame is given by $E_{\nu} = 2\gamma Q_{\rm EC}$, for fixed baseline
and number of useful decays, the ions with lower Q-values result in larger fluxes (because of the larger boost needed to achieve a given energy). The choice of ion is thus a balance
between Q-value and the available acceleration. From the ions presented in Tab. I, $^{148}$Dy and $^{154}$Er have very similar
characteristics; the bulk of all decays being an electron capture, half-lives $\sim$ 3 minutes and neutrino energies of $\sim$
2 MeV in the rest frame. The lower Q-value ion, $^{150}$Dy, will be taken in this paper for consistency with~\cite{catalina2} and
the higher $\gamma$'s required compensate for the lower electron capture branching ratio. The remaining 36~\% for $^{150}$Dy
is $\alpha$-decay and so does not source a primary neutrino background; the daughter has 100~\% $\alpha$-decay with a 74 year
half-life. The other ions in Tab. I have undesireable intensity and Q-value combinations. These higher Q-value ions
also have a large positron decay background and are more suitable for the hybrid machine introduced in [35].
For a $^{150}$Dy ion, the maximum boost attainable with an upgraded 1~TeV SPS~\cite{LHC_upgrade} is $\gamma_{\rm max}$ = 440\footnote{The maximum boost for a fully stripped $^{150}$Dy ion is in fact 
$\gamma=468$; however, it is necessary to leave several electrons bound to the nucleus to source the electron capture decay. If one leaves the 2 K-shell electrons and 2 in the L-shell, this reduces the maximum
boost to 440}  corresponding to
a laboratory frame energy of $E_{\nu}$ = 1.23 GeV. For a electron capture machine with no backgrounds the choice of
detector does not depend on the energy reconstruction capabilities as the event energy is determined by the boost. In
reality, some reconstruction might be necessary for the appearence events to reduce the atmospheric background,
especially if low production rates or issues with the acceleration force a large duty cycle. This point will be discussed
further in Sec.~\ref{S:bck}. Since the maximum laboratory frame energy is 1.23 GeV, baselines in excess of CERN-Canfranc
are unrealistic propositions. Therefore matter effects are small, but not negligible, and are not enough to achieve
competitive sensitivities to the mass hierarchy. Electron capture machines of this type are therefore `CP-violation
machines'.

   Although an electron capture machine is a variant on a Beta Beam, there are number of further technological
challenges that need to be overcome for such a facility to be realised;
\begin{enumerate}
\item Candidate ions have large proton numbers which bring problems with space charge in the early part of the
       accelerator chain.
\item Electrons need to be left on the ions for the electron capture channel to be available.
\item Decays rates are slow; typically several minutes. The optimal half-life for a Beta Beam type machine is $\mathcal{O}$(1 sec).
\end{enumerate}

   All these issues conspire to make the desired useful rate of $10^{18}$ ion decays per year extremely hard. To date,
no extensive R\&D has been carried out on the possible yearly rates for the rare-earth nuclei sought for electron
capture machines. However, code adapted from the baseline Beta Beam study has been used to simulate estimate
rates with the same vacuum and accelerator conditions~\cite{Fraser}. Assuming a round Gaussian beam of ions, the self-field 
incoherent tune shift was found to be large. Increased intensities in the acceleration only worsen the space
charge problem. This problem can be tempered by leaving extra electrons on the ions, thus reducing the charge;
however, such a strategy reduces the maximum boost attainable. This is not necessarily bad, as the results of the
simulation in Sec.~\ref{S:results} indicate that the maximum boost $\gamma$ = 440 may not be necessary. A possible strategy could
be to leave electrons bound to the ion such that the maximum boost is the sought boost. For instance, if one requires
a boost of $\gamma$ = 150, 44 electrons can be left bound to the Dy nucleus. The tune-shift has a charge-squared dependence, so we
decrease its effect by a factor of 9 with this action.

  This, however, brings with it a new problem. The vacuum conditions of the accelerator chain and decay ring are
such that the probability of electrons to be stripped from the ion is significant. Since this would alter the charge-to-mass 
ratio of the ion, the magnetic configuration of the accelerator or decay ring no longer matches that required and the
ion is lost. These losses can be thought of as an extra decay channel. The annual rate of neutrinos needs to be
modified~\cite{Fraser}:
\begin{equation}
R = \frac{I_{\rm in}l}{T_{\rm rep}}\frac{\lambda_{\rm EC}/\gamma}{\lambda_{\rm EC}/\gamma+\lambda_{\rm vac}}
                                                                          \left(1-e^{mT_{\rm rep}(\lambda_{EC}/\gamma +\lambda_{\rm vac})}\right)\:T_{\rm run}~.
\end{equation}
Here, $\lambda_{\rm EC}$ and $\lambda_{\rm vac}$ are the decay constants for the electron capture decay and vacuum losses respectively; $T_{\rm rep}$ is the
repetition period for fills in the decay ring; $I_{\rm in}$ the total number of ions injected into the decay ring for each fill; $l$
the livetime (the fraction of the decay ring length that sources the neutrino beam); m is the number of merges in the
decay ring and $T_{\rm run}$ the length of the experimental run in seconds. It is reasonable to assume $1/\lambda_{\rm vac} \sim 60$ secs~\cite{Fraser}.

  Ultimately though, the main problem in sourcing a competitive number of useful decays is the half-lives of the
ions. Sourcing a beam of mono-energetic neutrinos in this manner requires the use of ions with a fast Gamow-Teller
resonance. This requirement is very restrictive in terms of the choice of ions. Out of the candidates identified, the
fastest decays have weak EC-transitions; those with dominant resonances have much longer half-lives ($\sim$ 3 mins). To
realise an electron capture machine therefore will require a thorough search of the Gd region of the Segre chart for an ion possessing a
suitably fast resonant electron capture decay.

\section{The `energy' degeneracy}\label{S:ener}

It is well known that the analysis of data from a future long baseline facility suffers from the problem of degeneracies~\cite{deg1,deg2,deg3,deg4,deg5}; 
the asymmetry between neutrino and anti-neutrino probabilities, the unknown sign of $\Delta m^{2}_{31}$, and the
unknown octant of $\theta_{23}$ can all lead to multiple fits to experimental data. For binned data, the number of neutrino
(anti-neutrino) events in the ith neutrino (anti-neutrino) energy bin for the pair $(\bar{\theta}_{13}, \bar{\delta})$ is given by
\begin{equation}
N_{i}(\bar{\theta}_{13},\bar{\delta}) = \mathcal{N}_{T}\:t\:\int_{E_{i}}^{E_{i}+\Delta E}\: \epsilon(E_{\nu})\sigma_{\nu_{\mu}(\bar{\nu}_{\mu})}(E_{\nu})\:
P_{e\mu}^{\pm}(E_{\nu},\bar{\theta}_{13},\bar{\delta})\:\Phi_{\nu_{e}(\bar{\nu}_{e})}(E_{\nu})\:dE_{\nu} ~,
\end{equation}
where $\mathcal{N}_{T}$ is the number of targets in the detector, $t$ is the time of data taking, $\epsilon(E_{\nu})$ is the detector efficiency,
$\sigma(E_{\nu})$ is the interaction cross section, $\Phi(E_{\nu})$ is the beam spectrum and $\Delta E$ is the bin width.
Using the shorthand $\Delta_{ji} \equiv \Delta m^{2}_{ji}/(2E)$, the oscillation probability $P_{\nu_{e}\rightarrow \nu_{\mu}}\equiv P_{e\mu}$ can be expanded in the
small parameters $\bar{\theta}_{13}$, $\Delta_{21}/\Delta_{31}$, $\Delta_{21}/A$ and $\Delta_{21}L$~\cite{golden}, 
\begin{alignat}{1}\label{E:prob_exp}
P^\pm_{e \mu} (\bar \theta_{13}, \bar \delta)  &=  \sin^2 2 \bar\theta_{13} \, \sin^2 \bar \theta_{23} \left(\frac{\Delta_{31} }{B_\mp} \right)^2\sin^2 \left (\frac{B_\mp L}{2}\right) \notag \\ 
                                                                                &\quad + \mathcal{J}\, \frac{\Delta_{21}}{A} \frac{\Delta_{31} }{ B_\mp } \sin \left (\frac{A L }{ 2 } \right )\sin \left( \frac{ B_\mp L }{ 2 } \right )
                                                                                 \cos\left( \pm \bar \delta -\frac{\Delta_{31} L }{2}\right) \notag \\
                                                                                 &\quad \quad + \cos^2 \bar \theta_{23} \, \sin^2 2 \bar \theta_{12} \left(\frac{\Delta_{21}}{ A } \right )^2 \sin^2 \left(\frac{A L }{ 2}\right) ~,
\end{alignat}
where $\mathcal{J} = \cos\bar{\theta}_{13}\sin2\bar{\theta}_{12}\sin2\bar{\theta}_{23}\sin2\bar{\theta}_{13}$, the $\pm$ corresponds to neutrinos/anti-neutrinos and $B_\mp
\equiv A \mp\Delta_{31}$. Here we are using $A =
\sqrt{2}G_{F}\bar{n}_{e}(L)$ (the constant density approximation for
the index of refraction) where $\bar{n}_{e} =
1/L\int_{0}^{L}n_{e}(L')dL'$ is the average electron density and
$n_{e}(L)$ is the electron density along the baseline.

Labelling $N^{+}$ to be the number of neutrino events, and $N^{-}$ to be the number of anti-neutrino events; 4 pairs of equations can be solved
\begin{alignat}{1}
N^{\pm}(\bar{\theta}_{13},\bar{\delta},\vert \Delta m_{31}^{2}\vert, \theta_{23}) &= 
                                                      N^{\pm}(\theta_{13},\delta,\vert \Delta m_{31}^{2}\vert, \theta_{23})~, \\
N^{\pm}(\bar{\theta}_{13},\bar{\delta},\vert \Delta m_{31}^{2}\vert, \theta_{23}) &= 
                                                      N^{\pm}(\theta_{13},\delta,-\vert \Delta m_{31}^{2}\vert, \theta_{23})~, \\
N^{\pm}(\bar{\theta}_{13},\bar{\delta},\vert \Delta m_{31}^{2}\vert, \theta_{23}) &= 
                                                      N^{\pm}(\theta_{13},\delta,\vert \Delta m_{31}^{2}\vert, 90^{o}-\theta_{23})~, \\
N^{\pm}(\bar{\theta}_{13},\bar{\delta},\vert \Delta m_{31}^{2}\vert, \theta_{23}) &= 
                                                      N^{\pm}(\theta_{13},\delta,-\vert \Delta m_{31}^{2}\vert, 90^{o}-\theta_{23})~;
\end{alignat}
which in general lead to 8 solutions that can fit the data. In order, the first set of equations results in the `intrinsic
clone'; the second returns the `hierarchy clones'; the third gives the `octant' clones; and the fourth equation allows
for `mixed' clones. Solutions to the first set of equations are depicted graphically in Fig. 1 for $\sin^{2}2\theta_{13} = 10^{-2}$,
$\delta= 50^{o}$, $\theta_{23} = 45^{o}$ and $L = 650$ km. The black lines correspond to E = 1.3 GeV (first oscillation
maximum) and the red lines are for E = 1.8 GeV. Solid lines are for neutrinos and dashed lines are for anti-neutrinos.

\begin{figure}
\begin{center}
\includegraphics[width=9cm]{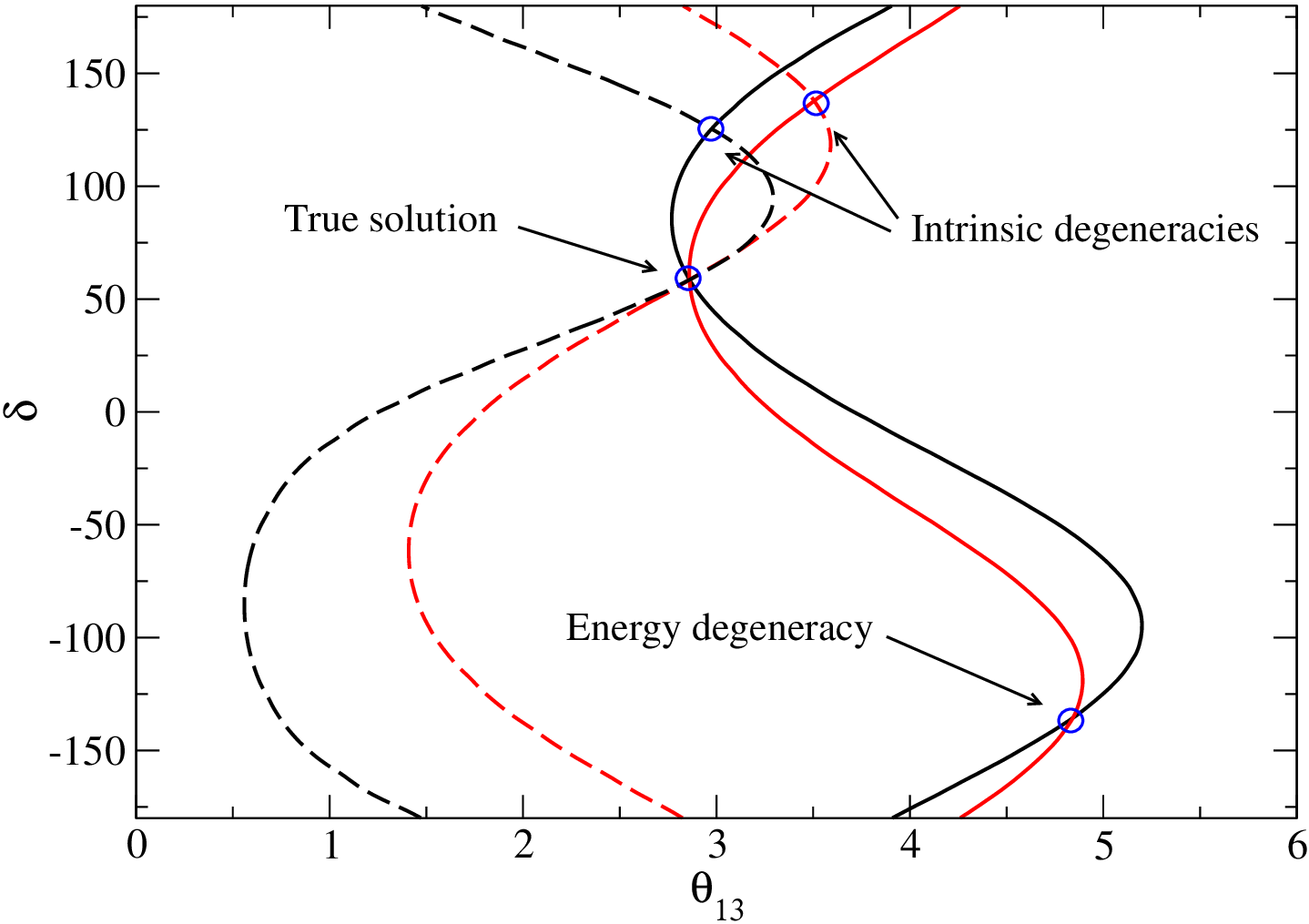}
\end{center}
\caption{Equi-probability curves for the CERN-Canfranc baseline (650 km) for $\sin\theta_{13}^{\rm tr} = 0.05$ and $\delta^{\rm tr} = 50^{o}$. The other 
oscillation parameters have been set to their current central values~\cite{STV08}. The black curves use $E_{\nu}=1.3$ GeV and the red curves use $E_{\nu}=1.8$ GeV.
Solid lines are for neutrinos, dashed lines are for anti-neutrinos.}\label{Fi:enerdeg}
\end{figure}

   From Fig.~\ref{Fi:enerdeg}, there is another type of degeneracy not discussed in the literature. Consider the first
set of equations above but now also for energies $E_{1}$ and $E_{2}$ :
\begin{equation}\label{E:ener_eq}
N^{\pm}_{1,2}(\bar{\theta}_{13},\bar{\delta},\vert \Delta m_{31}^{2}\vert, \theta_{23}) = 
                                                      N^{\pm}_{1,2}(\theta_{13},\delta,\vert \Delta m_{31}^{2}\vert, \theta_{23})~.
\end{equation}
Attempting to find a solution to these equations will return only the true solution since the location of the intrinsic
clone is energy dependent. This is the main strategy in resolving the intrinsic degeneracy. Suppose, one does not
have an anti-neutrino channel. This set reduces from 4 to 2 equations, and in general will possess a clone solution.
This can been seen as the intersection of the red and black solid curves at ($\theta_{13},\delta) = (4.8^{o},-140^{o})$. The location
of this degeneracy will be different for neutrino and anti-neutrino runs. Therefore the inclusion of both resolves
the degeneracy and hence it is not present in most setups discussed in the literature. For single ion Beta Beams
and electron capture studies~\cite{singleion,ECBB, catalina1, catalina2}, this degeneracy needs to be resolved by the experiment or constrained to
values of $\theta_{13}$ larger than near future experimental limits. This degeneracy shall hereafter be referred to as the `energy
degeneracy' to distinguish it from the usual intrinsic degeneracy which has a different origin.

Ideally, we would want the energy degeneracy to not be present in the data. To find a condition for this to be so, first rewrite the probability (Eq.~\ref{E:prob_exp}) as
\begin{equation}
P_{e\mu}= I_{1}\sin^{2}2\bar{\theta}_{13}+I_{2}\sin2\bar{\theta}_{13}\cos\bar{\delta}+I_{3}\sin2\bar{\theta}_{13}\sin\bar{\delta}+I_{4}
\end{equation}
so that all the non-essential constants are tied up in the $I_{i}$. As a first step, we attempt to solve Eq.~\ref{E:ener_eq} for mono-energetic neutrinos only with energies
$E_{1}$ and $E_{2}$. Labelling the respective coefficients as $I_{i}^{1}$ and $I_{i}^{2}$ we obtain the relation
\begin{equation}\label{E:eq_solve}
\left[\frac{I_{1}^{1}}{I_{3}^{1}}-\frac{I_{1}^{2}}{I_{3}^{2}}\right](\sin^{2}2\theta_{13}-\sin^{2}2\bar{\theta}_{13})
+\left[\frac{I_{2}^{1}}{I_{3}^{1}}-\frac{I_{2}^{2}}{I_{3}^{2}}\right](\sin2\theta_{13}\cos\theta_{13}\cos\delta-\sin2\bar{\theta}_{13}\cos\bar{\theta}_{13}\cos\bar{\delta}) = 0~.
\end{equation}
For the energy degeneracy to be resolved, we must have $\theta_{13} = \bar{\theta}_{13}$ which is true if either
\begin{enumerate}
\item $I_{2}^{1}/I_{3}^{1}=I_{2}^{2}/I_{3}^{2}$, i.e. $\Delta m_{31}^{2}/4E_{1}$ and $\Delta m_{31}^{2}/4E_{2}$ differ by $\pi$;
\item $I_{1}^{1}/I_{3}^{1}=I_{1}^{2}/I_{3}^{2}$.
\end{enumerate}
In general $I_{1}^{1}/I_{3}^{1}\ne I_{1}^{2}/I_{3}^{2}$ so we are left with the first condition. The set of equations~(\ref{E:ener_eq}) can accommodate the solution
\begin{equation}
\delta=\pi-\bar{\delta}-2\tan^{-1}\left(\frac{I_{2}^{1}}{I_{3}^{1}}\right)~,
\end{equation}
in addition to the trivial solution $\delta = \bar{\delta}$ for the case $\theta_{13}=\bar{\theta}_{13}$. Therefore the energy degeneracy for a pair of neutrino 
energies is only completely resolved if condition 1 above and 
\begin{equation}
\bar{\delta}=\frac{\pi}{2}-\tan^{-1}\left(\frac{I_{2}^{1}}{I_{3}^{1}}\right)
\end{equation}
hold. Therefore, the energy degeneracy is present in general for two beams of mono-energetic neutrinos; however, this
is not necessarily a nuisance. In section~\ref{S:results}, it will be found that a combination of mono-energetic neutrino beams
placed almost on first oscillation maximum and on second oscillation maximum provides some of the largest CP-violation 
sensitivity coverage. In such a case, the CP-even part of the probability vanishes and the energy degeneracy
is located at $\delta = \pi - \bar{\delta}$. Since this change will leave the probability invariant, the degenerate region will be the same
strength and will be symmetrically placed about the $\delta = \pi/2$ or $-\pi/2$ lines. The two regions will always be either be
CP-conserving or CP-violating at the same time.

   In~\cite{ECBB,singleion} only neutrinos were used. In those studies, the strategy was to exploit the energy dependence of the
oscillation signal to break degeneracy and push for a good physics reach; specifically though the combination of bins
centred around first and second oscillation maximum. The above argument says that the combination of the maxima
is insufficient to completely break the energy degeneracy, but the $\theta_{13}$-part of the degeneracy is broken. The reason
why the degeneracy was only present for very small values of $\sin^{2}2\theta_{13}$ in these studies was because the data was
binned. If one thinks of the data set as predominantly pairs of bins separated by $\delta \Delta m_{31}^{2}/4E = \pi$, the location of the
energy degeneracy is different for each pair and only the true solution is statistically significant. Or more simply, from Eq.~\ref{E:eq_solve},
the combination of multiple energies completely breaks the degeneracy as its location is energy dependent.
In this paper, only two electron capture boosts will be used and so this degeneracy is in general present.

\section{A dual boost electron capture machine}\label{S:analysis}

In this section, a summary of the simulations carried out will be given. First, the two Lorentz boosts will be varied independently to identify pairs of interest.
To begin with the number of useful decays per year will be fixed at $N_{\rm ions} = 10^{18}$, irrespective of the boost, and
no atmospheric backgrounds will be included. Since both these parameters are critical for the projective CP-sensitivity,
they will be introduced shortly once pairs have been identified for further study. The neutrino oscillation parameters have
been set to the current values~\cite{STV08}. For all simulations, the Gaussian form of the $\chi^{2}$ function is used:
\begin{equation}
\chi^{2}= \sum_{i}\frac{(n_{i}-\zeta_{i})^{2}}{\zeta_{i}+(f_{\rm sys}\cdot\zeta_{i})^{2}}~,
\end{equation}
where $n_{i}$ is the `true' event rate for boost $\gamma_{i}$ and $\zeta_{i}$ is the corresponding `test' event rate. Here $f_{\rm sys} = 2~\%$ is an overall
normalisation error. An intrinsic background of 0.1~\% of the unoscillated flux is included to take into account neutral
current pion production and electrons misidentified as muons. Inclusion of atmospheric neutrinos will be discussed in
section~\ref{S:bck}. Throughout, a 440 kton Water \v{C}erenkov is used with no energy reconstruction. This allows for all charge
current events to be taken as signal, as oppose to just the quasi-elastic events. The downside of this approach is that
all beam and atmospheric backgrounds that pass the cuts will be included as signal regardless of their energy.

The sensitivity to CP-violation of an electron capture machine as a function of two boosts, $\gamma_{1}$ and $\gamma_{2}$, and the run
time fraction, $f$, will be explored. For a given boost pair $(\gamma_{1}, \gamma_{2})$, number of targets $\mathcal{N}_{T}$, and run time fraction $f$, the
simulated event rate is 
\begin{equation}
n_{\gamma_{i}}=\mathcal{N}_{T}\:\eta_{i}\:t\int_{0}^{\infty}\Phi(E_{\nu})\:\sigma(\nu_{\mu})P_{\nu_{e}\rightarrow\nu_{\mu}}\:\delta(E-E_{\nu})\:dE~,
\end{equation}
where $\eta_{1}=f$, $\eta_{2}=1-f$ and t is the total run time. Although, the CERN-Canfranc baseline is relatively short, the full oscillation probability
was numerically simulated including matter effects.

For an electron capture machine, one has complete freedom (within technological bounds) to choose the Lorentz boosts of the ion. Clearly, while some choices 
of boosts will return good sensitivity to CP-violation for given ranges of $\sin^{2}2\theta_{13}$, others will result in poor sensitivity. Further, although the boost choice 
in~\cite{catalina2} returned a very competitive minimum $\sin^{2}2\theta_{13}$, the sensitivity region was asymmetric in $\delta\rightarrow -\delta$ and there existed
degeneracy for $\sin^{2}2\theta_{13}\sim 10^{-2}$ and $\delta<0$. Generating and then preparing CP-sensitivity plots for many pairs $(\gamma_{1},\gamma_{2})$
to compare visually is both computationally intensive and a cumbersome undertaking. The task at hand, therefore, is to adopt a measure to select boost pairs that return
a good overall CP-violation reach or those that are of interest for particular ranges of $\sin^{2}2\theta_{13}$. One approach would be to determine the minimum 
$\sin^{2}2\theta_{13}$ at a given confidence level for a particular experimental parameterisation; however, for an electron capture machine, this approach is not
desirable for two reasons. As mentioned previously, in the previous study~\cite{catalina2}, the CP-sensitivity region was asymmetric with considerably
less sensitivity for $\delta > 0$. Secondly, there were regions, especially for $\sin^{2}2\theta_{13}\sim 10^{-2}$ and $\delta < 0^{o}$, where degeneracy
persisted. Choosing a particular value for $\delta$ then searching for the pair with the minimal $\sin^{2}2\theta_{13}$ will not take into account these
features. It could be the case that a boost pairing with sensitivity to CP-violation down to very small $\sin^{2}2\theta_{13}$ has very poor sensitivity at larger 
$\sin^{2}2\theta_{13}$.

In this article, a measure referred to as the `CP-coverage' ($\delta_{\rm cov}$) is adopted. This measure is essentially an approximation to an `integrated CP-fraction', and is described 
below. For a particular pair $(\gamma_{1}, \gamma_{2}$), the CP-coverage is determined in the following manner
\begin{enumerate}
\item For $\sin^{2}2\theta_{13}\equiv s_{2;13}\in \left[s_{2;13}^{\rm min},s_{2;13}^{\rm max}\right]$, CP-violation is tested; for each pair $(s_{2;13},\delta)$ on a grid
for the $s_{2;13}$ range defined previously and $\delta\in\left[-180^{o},180^{0}\right]$, a fit to both $\delta = 0^{o}$ and $\delta =180^{o}$ is attempted. Solutions
for the opposite ${\rm sign}(\Delta m_{31}^{2})$ are also tested. The minimum $\chi^{2}$ ($\chi^{2}_{\rm min}$) so calculated is taken.
\item In this article, 2 degrees of freedom and 99~\% confidence levels are adopted. If $\chi^{2}_{\rm min}>9.21$ for a particular pair $(s_{2;13},\delta)$, then 1 is added to a tally.
\item The fraction of grid points for which CP-conservation can be ruled out is the `CP-coverage' for chosen range 
$s_{2;13}\in \left[s_{2;13}^{\rm min}, s_{2;13}^{\rm max}\right]$.
\end{enumerate}
This procedure is repeated for a grid of $(\gamma_{1},\gamma_{2})$ pairs. The CP-coverage is a discrete approximation to the integrated CP-fraction. Let the 
CP-fraction, $\delta_{\mathcal{F}}=\delta_{\mathcal{F}}(s_{2;13})$ be the fraction of $\delta$ for a given $\sin^{2}2\theta_{13}$ for which CP-conservation can
be ruled out. Then
\begin{equation}
\delta_{\rm cov} \longrightarrow \frac{1}{\mathcal{A}}\int_{s_{2;13}^{\rm min}}^{s_{2;13}^{\rm max}} \delta_{\mathcal{F}}\:ds_{2;13}
\end{equation}
as the number of grid points taken in the range $s_{2;13}\in \left[s_{2;13}^{\rm min},s_{2;13}^{\rm max}\right]$ and $\delta\in\left[-180^{o},180^{0}\right]$ tends to 
infinity. Here, $\mathcal{A}$ is the area of the region defined by $\left[s_{2;13}^{\rm min},s_{2;13}^{\rm max}\right]$ and $\delta\in\left[-180^{o},180^{0}\right]$. The CP-coverage is thus an approximation to the fractional area of the region in ($s_{2;13},\delta)$ space for which CP-conservation can be ruled out.

In the next section, CP-coverage will be used in two stages: firstly for $10^{-5}<\sin^{2}2\theta_{13}<10^{-1}$; and then for each individual decade. Choosing a large
range of integration allows one to identify boost pairs that may have the desirable CP-features; namely discovery reach down to $\sin^{2}2\theta_{13}\sim 10^{-4}$ and minimal
degeneracy for larger values. However, this choice is by no means perfect since if one requires a machine configured for $\sin^{2}2\theta_{13}\in [10^{-3},10^{-2}]$, say, the 
information for $\sin^{2}2\theta_{13}<10^{-3}$ can distort the relevance of the result. Specifically, it is quite possible that a facility with a good minimal $\sin^{2}2\theta_{13}$ may
have a poor return with regards CP-violation for large ranges of $\sin^{2}2\theta_{13}$. Therefore, to search for further pairs of interest, it is necessary to repeat the analysis for
smaller $\sin^{2}2\theta_{13}$ intervals.

\section{Results}\label{S:results}

\begin{figure}
\begin{center}
\includegraphics[width=10cm]{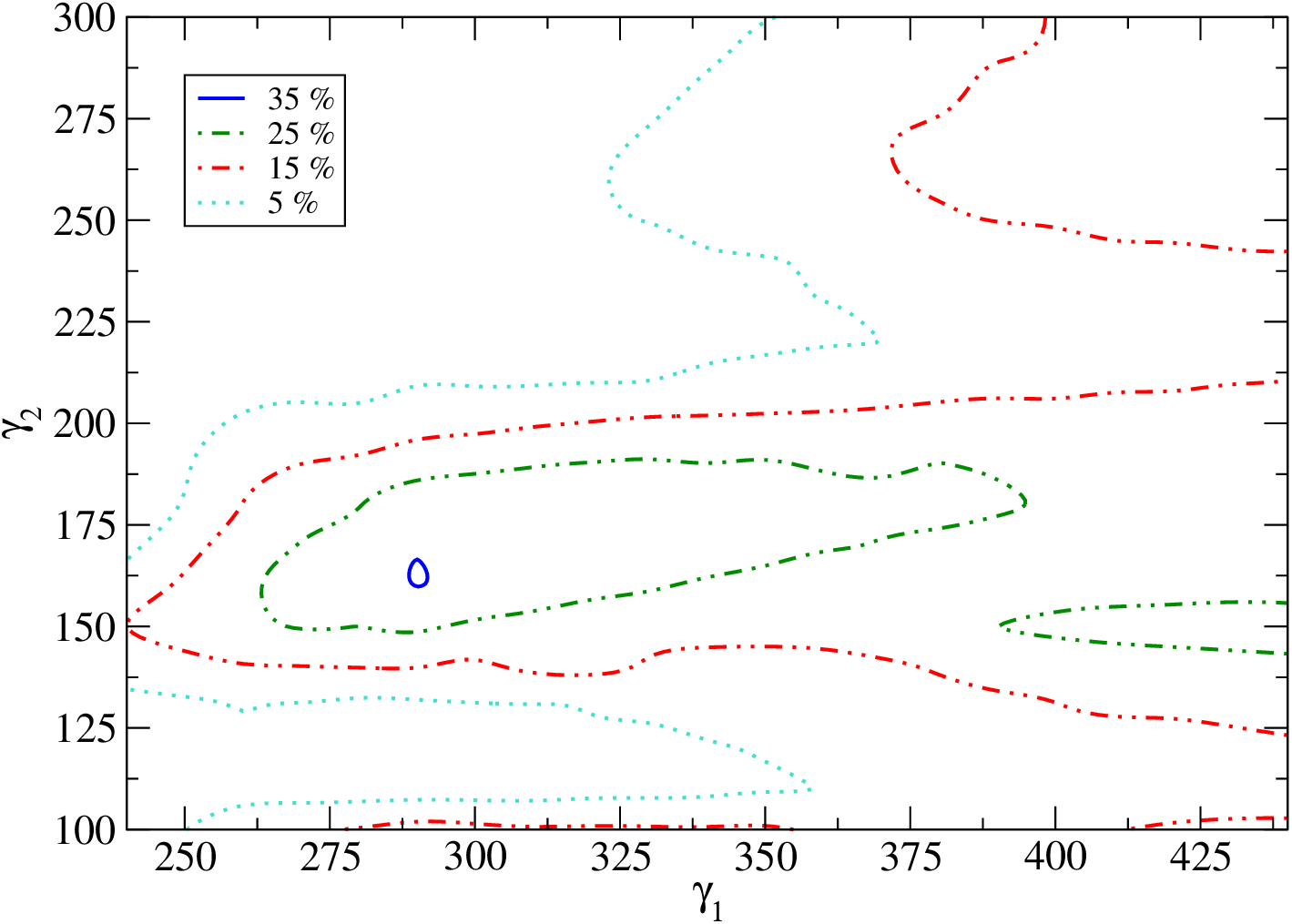}
\end{center}
\caption{For $f=0.3$, the CP-coverage as a function of $\gamma_{1}$ and $\gamma_{2}$.}\label{Fi:boost_pairs}
\end{figure}

   In Fig.~\ref{Fi:boost_pairs}, the 5~\%, 15~\%, 25~\% and 35~\% coverage contours are presented for the case $f = 0.3$. There are two distinct
regions in the \gp~plane; the first corresponds to the placement of the neutrino flux at the highest energy allowed in
combination with a flux at second oscillation maximum. This combination corresponds to the naive expectation that
good sensitivity should result from combining first and second oscillation maximum. However, there is a (additional)
large region of \gp~ space in which sizeable coverage can be achieved. Specifically, for a high boost, $260 < \gamma_{1} < 400$
and a narrower range of low boosts, $150 < \gamma_{2} < 185$; there is at least 25~\% coverage of the \td~ plane. This rises
to 35~\% for \gp~ $\sim (280, 160)$. This result is repeated for the cases $f = 0.4$ and $f = 0.5$ with little variation.
The small region with 35~\% coverage at (280, 160) is the best CP-coverage of the run time fractions and boost pairs
studied. The $f=0.3$ case runs at $\gamma=150$ for the longest period, and hence is the easiest technological option. This case will be considered henceforth.

\begin{figure}
\begin{center}
\includegraphics[width=7cm]{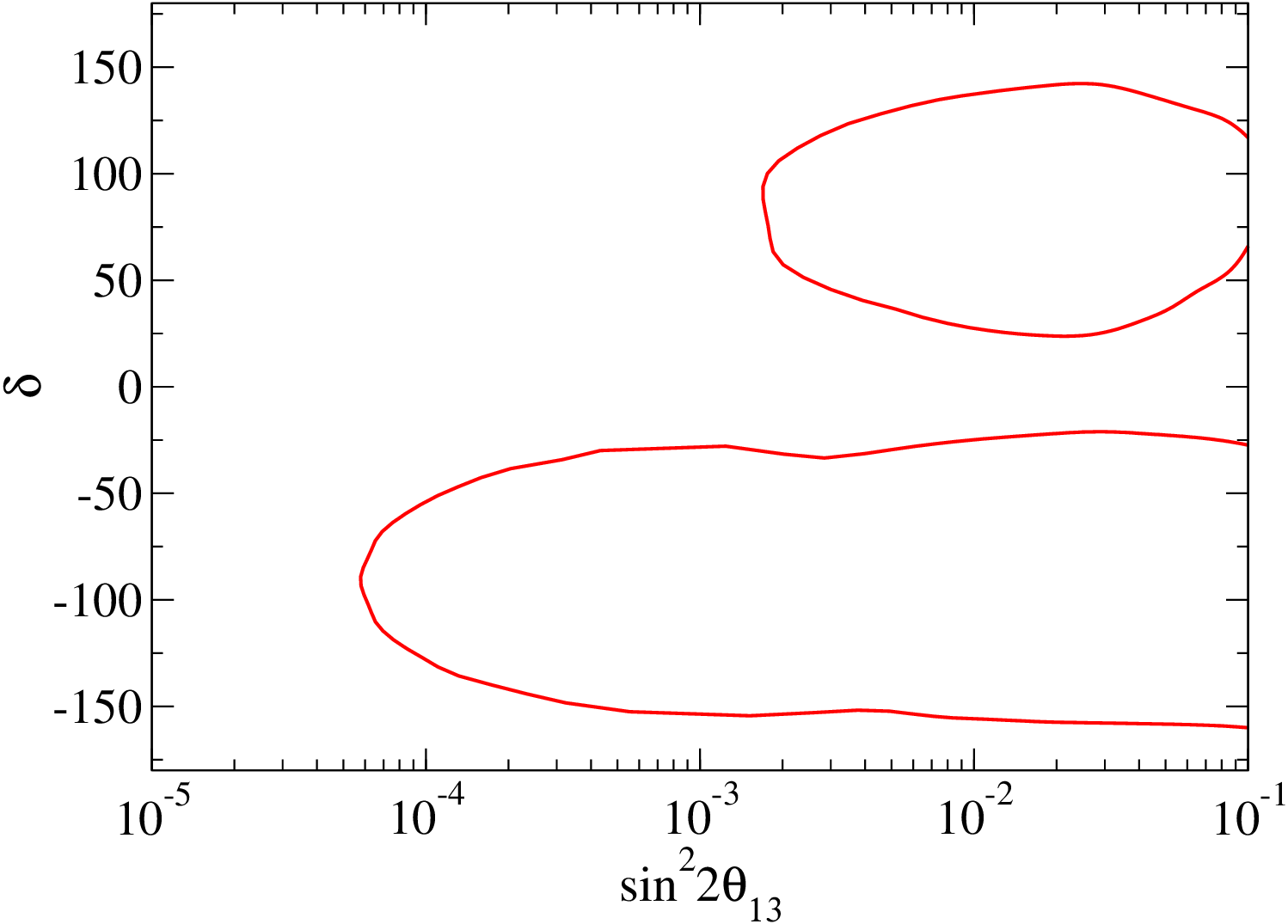}
\includegraphics[width=7cm]{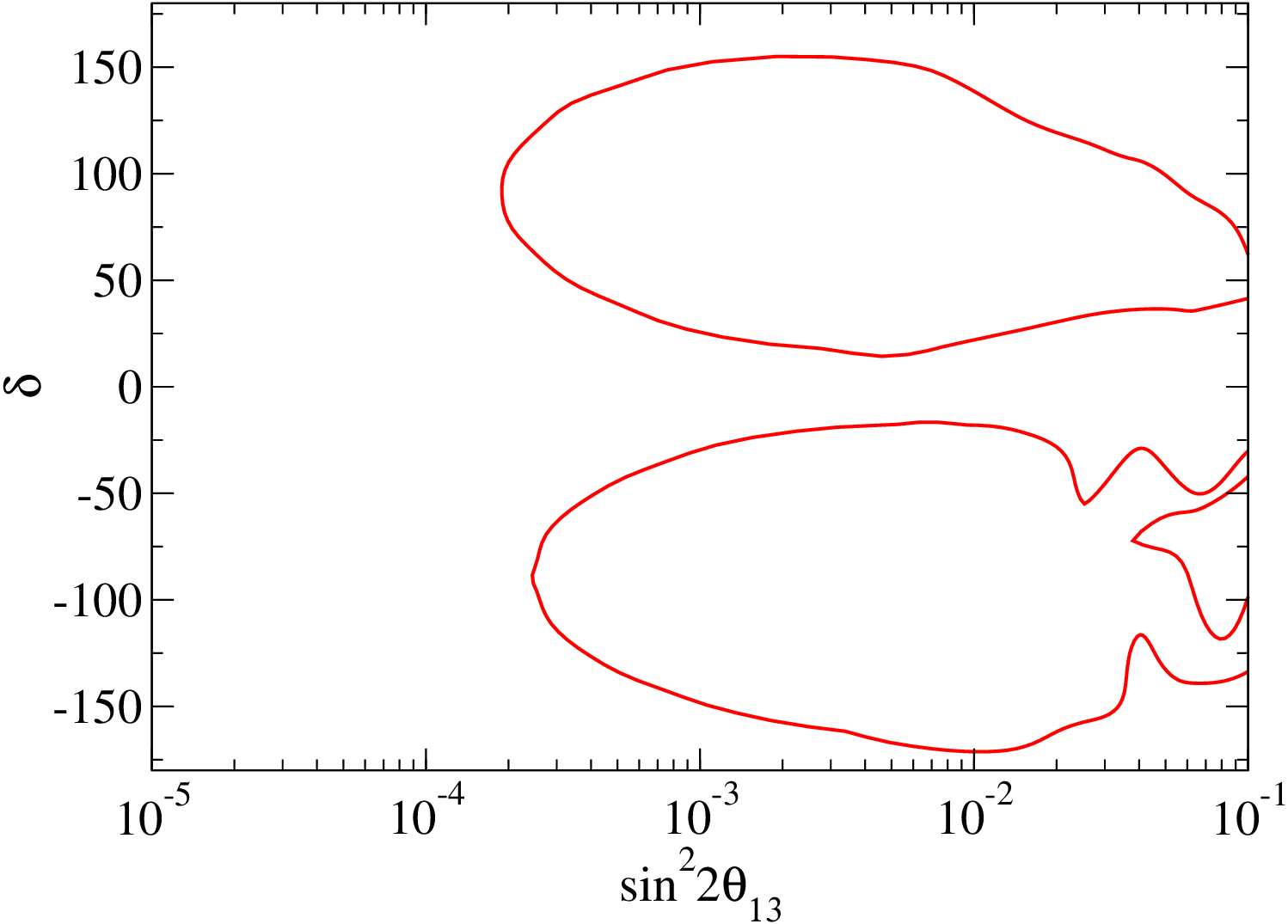}
\end{center}
\caption{CP-violation sensitivity at 99~\% for the pairs \gp $=(440,150)$ (left) and \gp $=(280,160)$ (right). }\label{Fi:CP}
\end{figure}

In Fig.~\ref{Fi:CP} the CP-sensitivity plots are shown for the two pairs (440, 150) and (280, 160). These correspond to
the centre of the `naive' choice of boosts and maximum coverage region, respectively. It is seen that the first option
produces an asymmetrical sensitivity region, the best sensitivity for $\delta < 0^{o}$; any degeneracy is resolved in larger
regions of parameter space. However, the minimal \s2 is larger for $\delta > 0^{o}$ and this choice of boosts. For $\delta < 0^{o}$,
the lack of sensitivity around \s2 $\sim 10^{-2}$ that was present in the original study~\cite{catalina2} is not present and the
minimal \s2 for which CP-conservation can be ruled out at 99~\% is also slightly smaller. 

\begin{figure}
\begin{center}
\includegraphics[width=6.5cm]{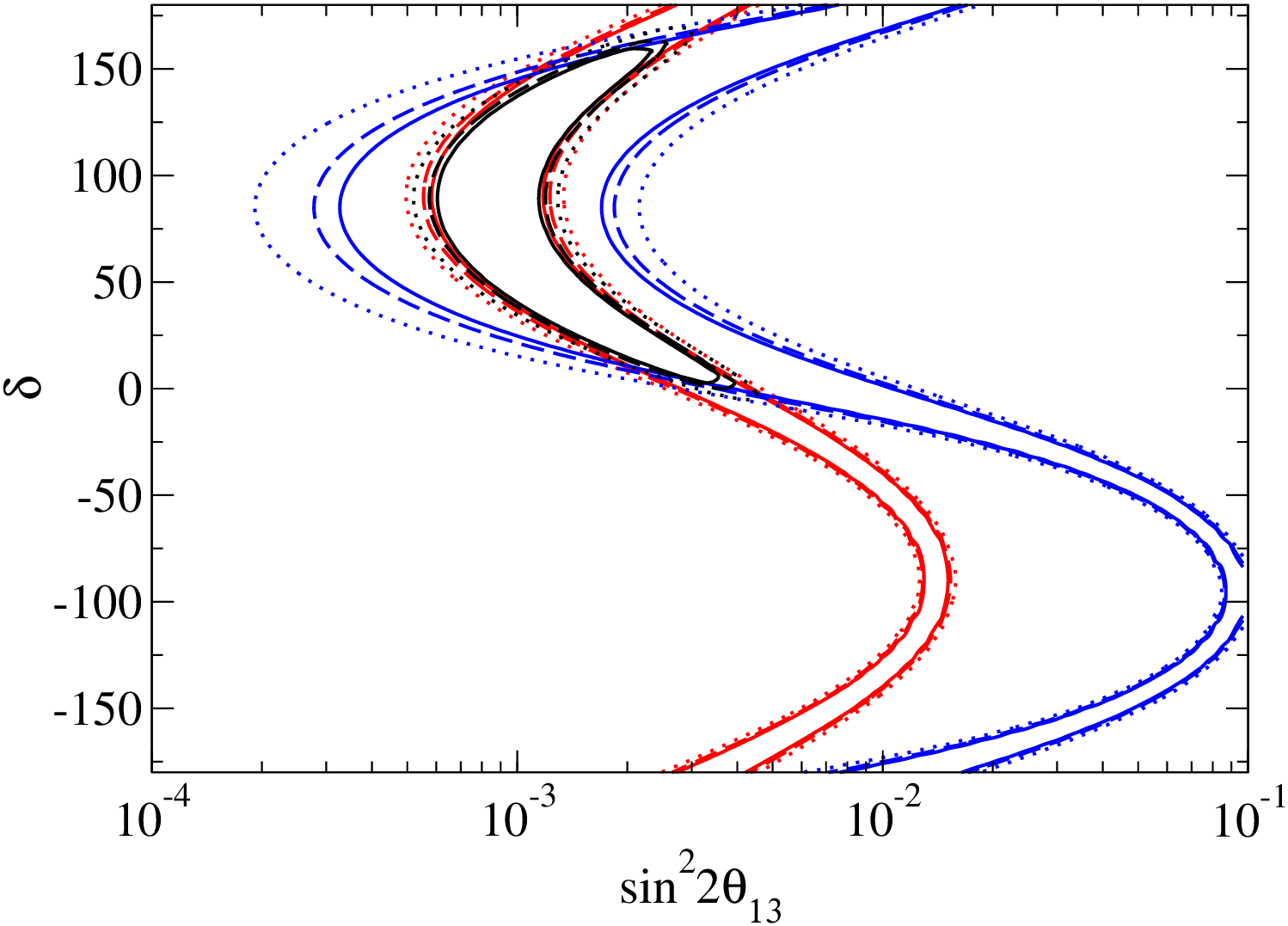}
\includegraphics[width=6.5cm]{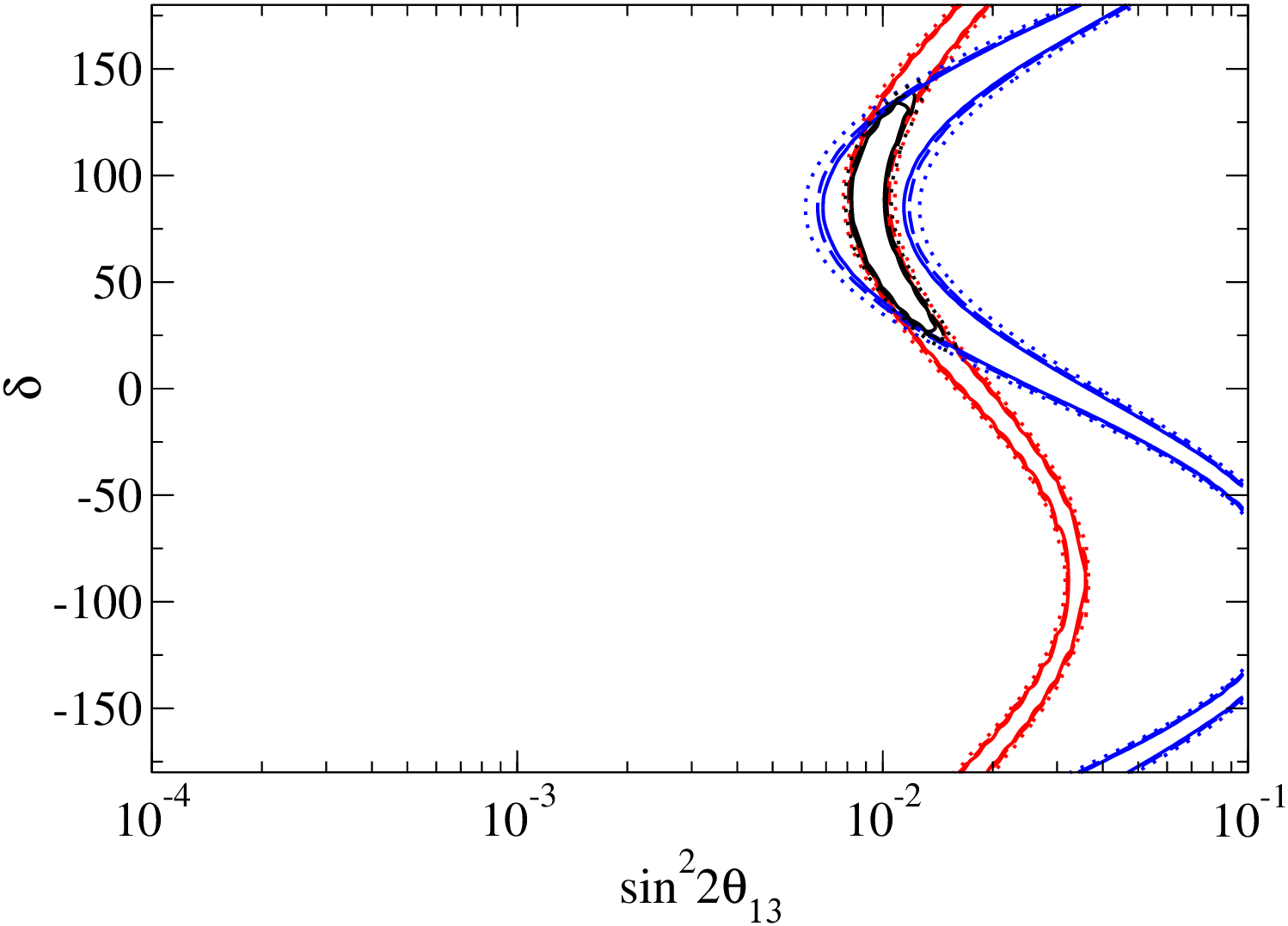}\\
\includegraphics[width=6.5cm]{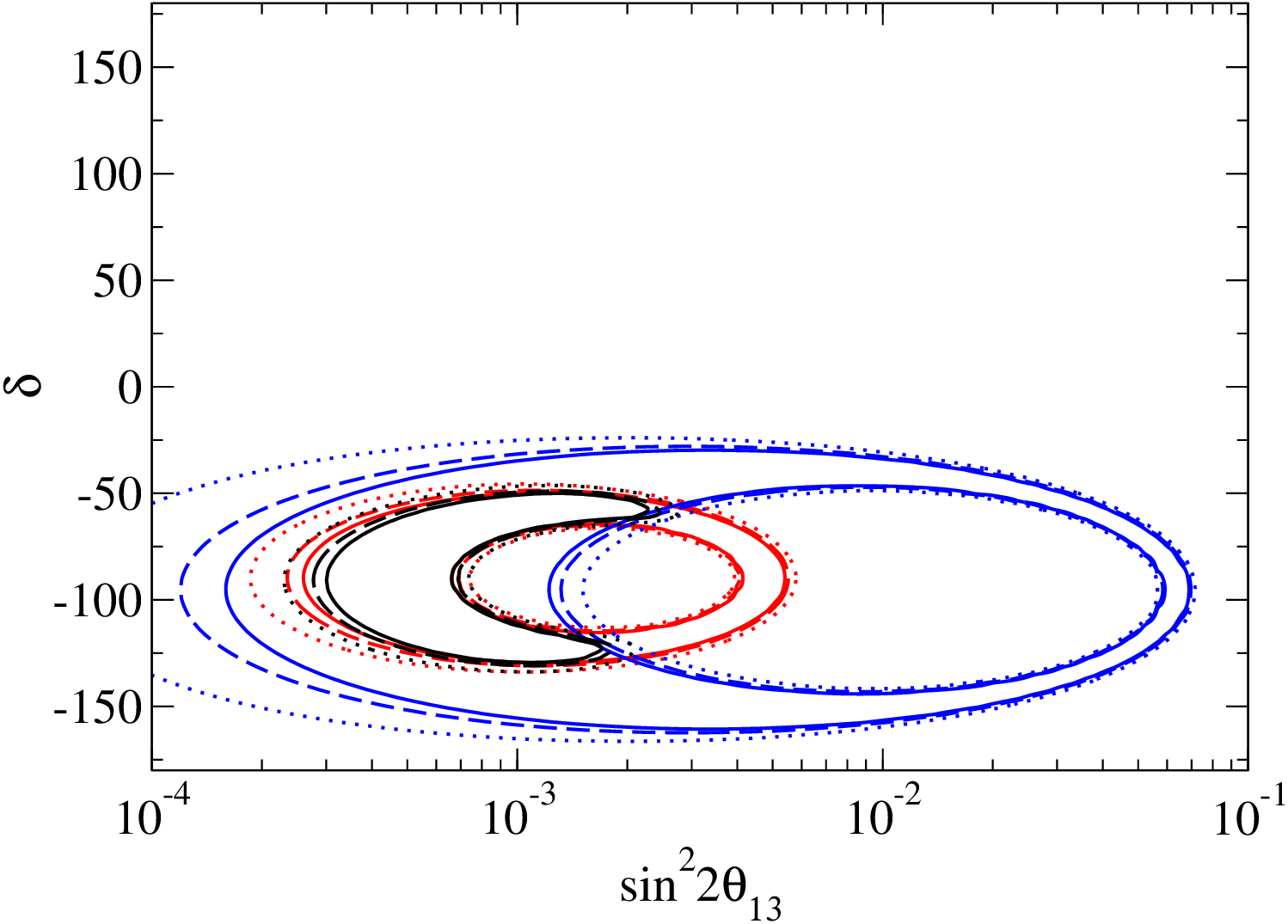}
\includegraphics[width=6.5cm]{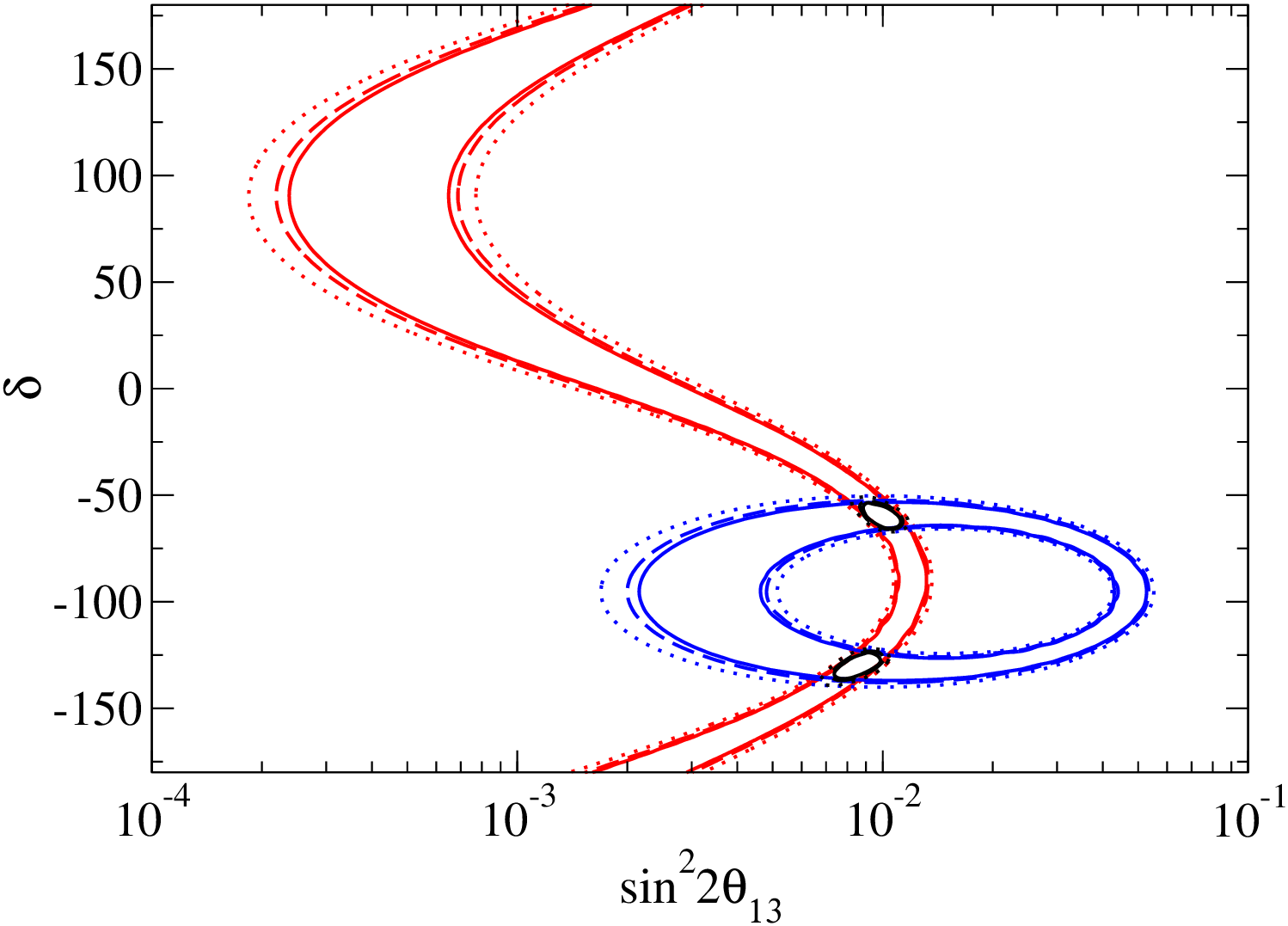}\\
\end{center}
\caption{90\%, 95\% and 99\% C.L. 2 parameter fits for $(\gamma_{1},\gamma_{2})=(440,150)$ for the CERN-Canfranc baseline 
(650 km). Plots have been produced on the assumption that $\sin^{2}2\bar{\theta}_{13}=10^{-3}$ (left), $10^{-2}$ 
(right), $\bar{\delta}=60^{o}$ (top) and $\bar{\delta}=-60^{o}$ (bottom). The blues curves correspond to $\gamma=440$,
the red to $\gamma=150$ and the black to the overall sensitivity. Normal mass ordering has been assumed. }
\label{Fi:sensplots_asym}
\end{figure}

\begin{figure}
\begin{center}
\includegraphics[width=6.5cm]{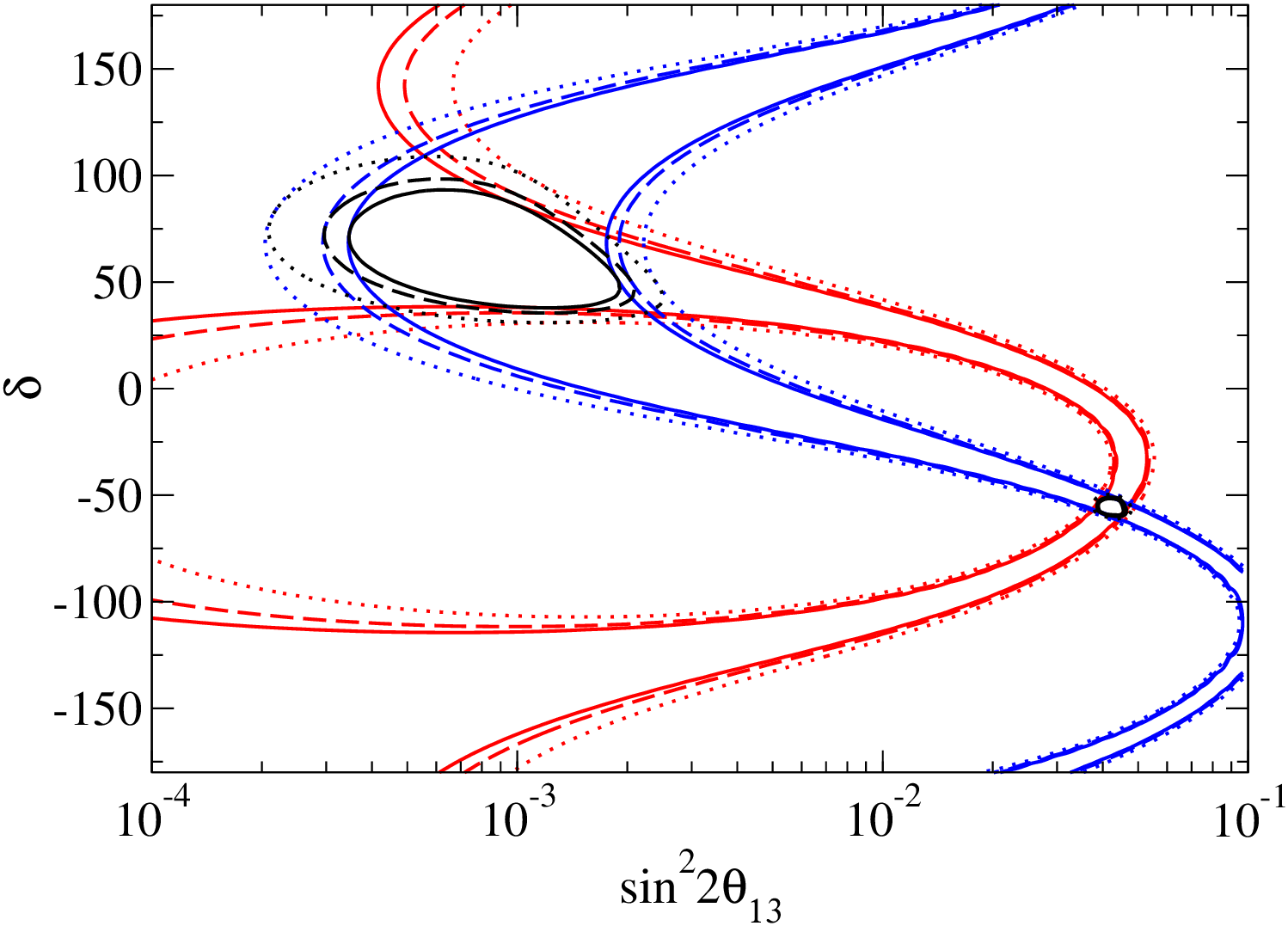}
\includegraphics[width=6.5cm]{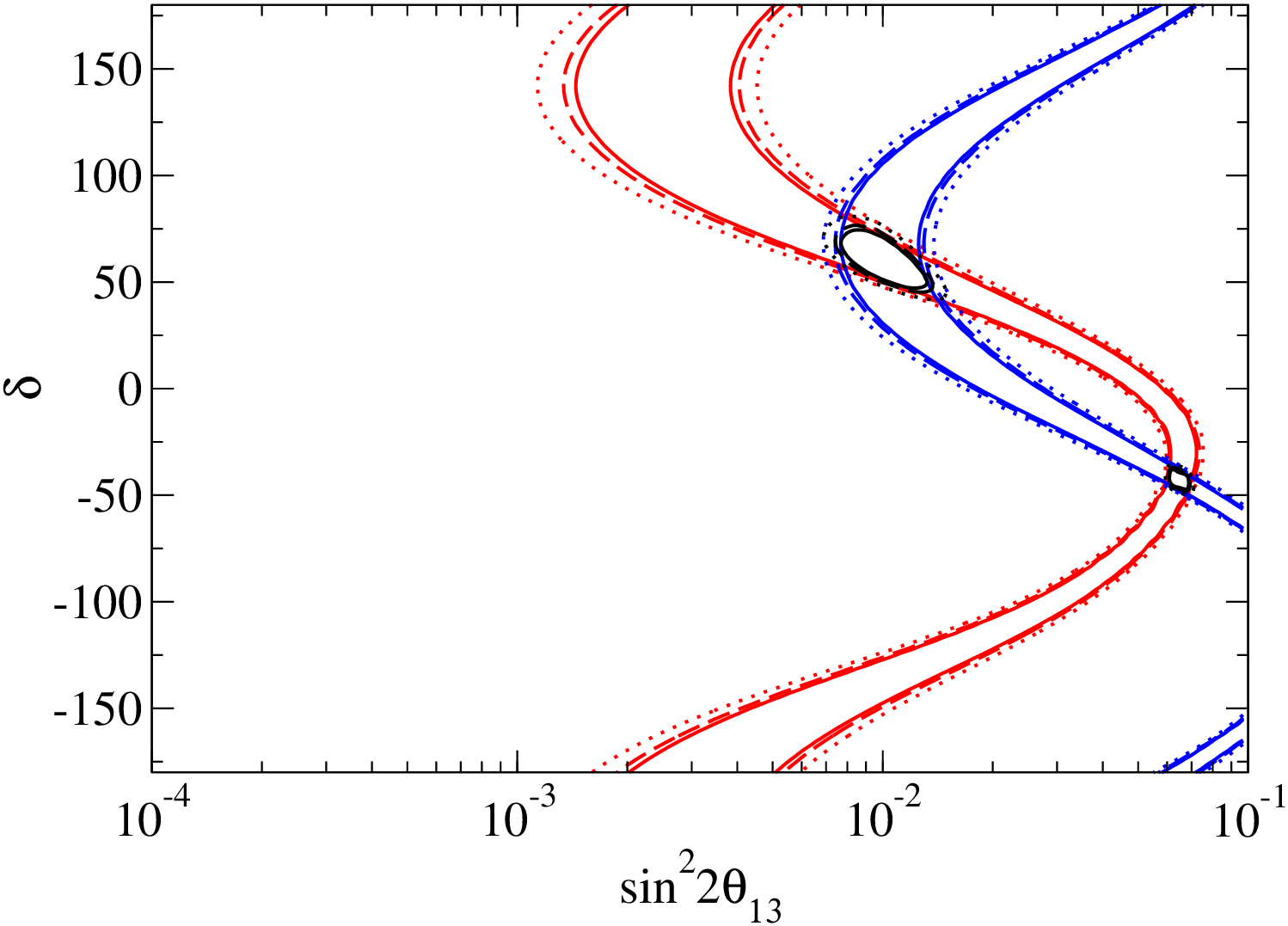}\\
\includegraphics[width=6.5cm]{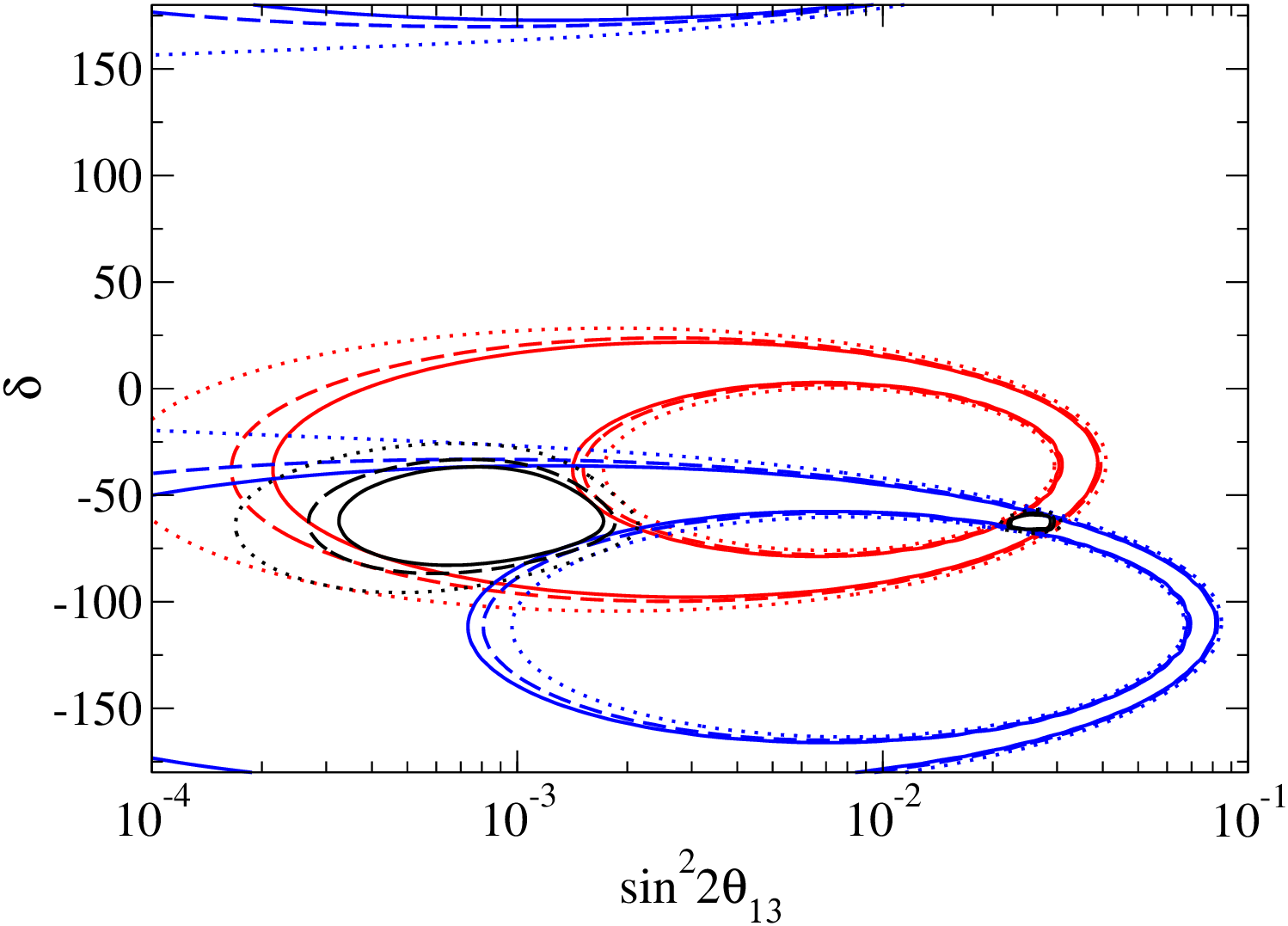}
\includegraphics[width=6.5cm]{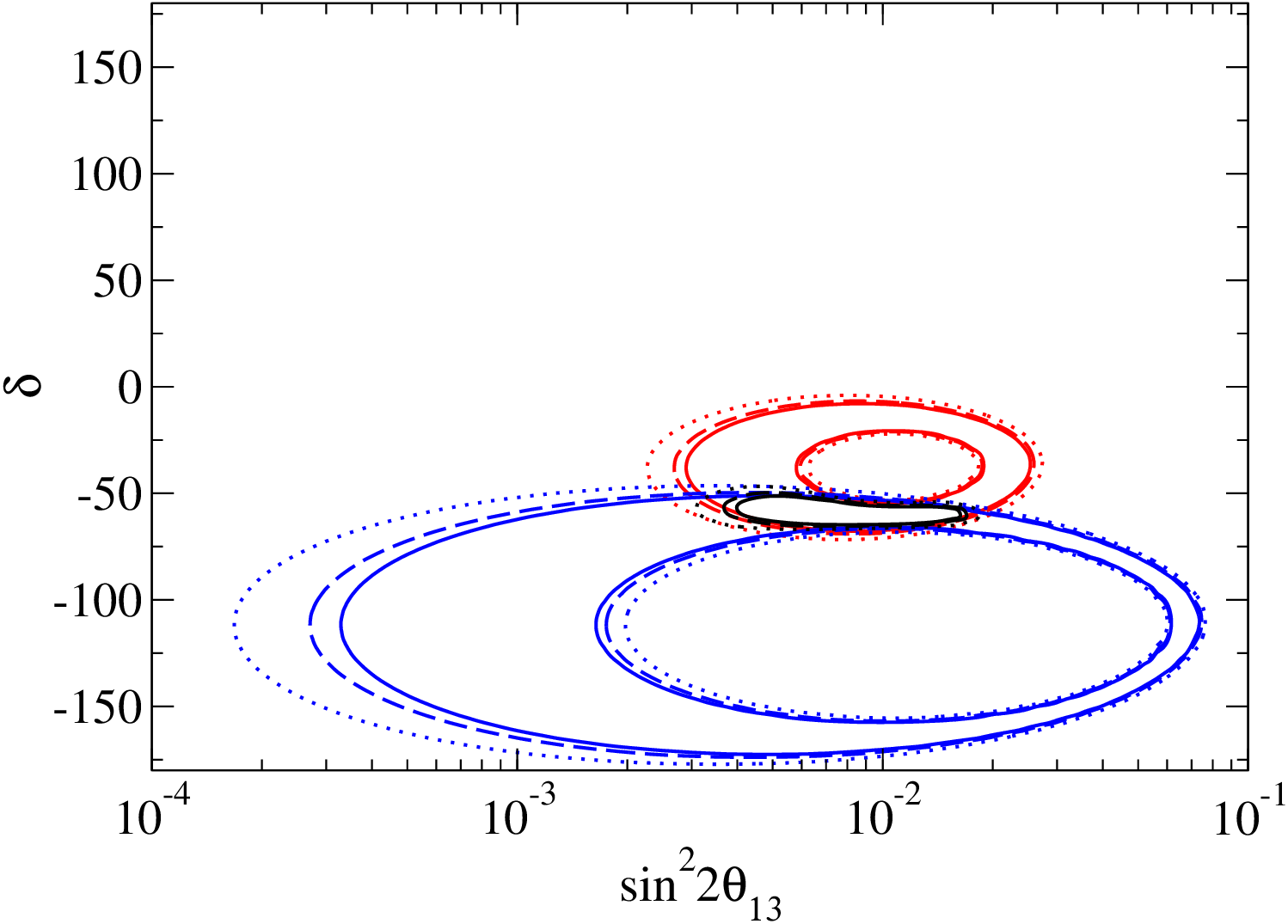}\\
\end{center}
\caption{90\%, 95\% and 99\% C.L. 2 parameter fits for $(\gamma_{1},\gamma_{2})=(280,160)$ for the CERN-Canfranc baseline 
(650 km). Plots have been produced on the assumption that $\sin^{2}2\bar{\theta}_{13}=10^{-3}$ (left), $10^{-2}$ 
(right), $\bar{\delta}=60^{o}$ (top) and $\bar{\delta}=-60^{o}$ (bottom). The blues curves correspond to $\gamma=280$,
the red to $\gamma=160$ and the black to the overall sensitivity. Normal mass ordering has been assumed.
}\label{Fi:sensplots_sym}
\end{figure}

   From the right-hand panel of Fig. 3, it is seen that the combination of two lower boosts results in a more symmetrical
sensitivity region in the \td~plane. For large values of \s2, the effects of degeneracies are still present, especially
for $\delta < 0^{o}$. This is of little concern, however, since measurement of \s2 $> 10^{-2}$ will be explored with the next
generation reactor and accelerator long baseline experiments. The minimum \s2 for which CP-violation can be
established is now a factor of 4 larger. This feature is a result of a large region of equivalent solutions being present
(at 99~\% confidence level) at small values of \s2 (smaller boosts imply smaller event rates which weaken the
sensitivity). By choosing two relatively small boosts, the setup has been configured to explore the interference features 
of the appearance probability. In doing so, there is now a poorer resolution on \t13 which comes predominantly from
the atmospheric features where there is little degeneracy between \t13 and \d.

In Fig.~\ref{Fi:sensplots_asym}, 90~\%, 95~\%
and 99~\% C.L. 2 parameter fits for the pair (440,150) are presented. The overall sensitivity and the contributions from
each boost are shown and have been computed on the assumption of normal mass ordering. The four true value pairs
$(\sin^{2}2\bar{\theta}_{13},\bar{\delta}) = (10^{-3}, 60^{o})$, $(10^{-3},-60^{o})$, $(10^{-2}, 60^{o})$ and $(10^{-2},-60^{o})$ are shown. 
The choice of boosts corresponds to placing the electron capture flux on second oscillation maximum and, approximately, first oscillation maximum.
The quantities $\Delta m_{31}^{2}L/4E_{1}$ and $\Delta m_{31}^{2}L/4E_{2}$ are therefore $\pi$ out of phase with each other. The sinusoidal shape in
solely the cause of interference effects; the solar and atmospheric features of the appearance probability have no $\delta$
dependence. In the lower right panel, the energy degeneracy is seen to be explicitly present. For the choices $\bar{\delta} = 60^{o}$
and $\bar{\delta} = -60^{o}$, one would expect to see degeneracy at the same $\theta_{13}$ and $\delta = 120^{o}$ or $-120^{o}$. These degeneracies are in
fact present but the large sensitivity regions for the $\gamma = 150$ run have caused them to merge into one large sensitivity
region.

\begin{figure}
\begin{center}
\includegraphics[width=14cm]{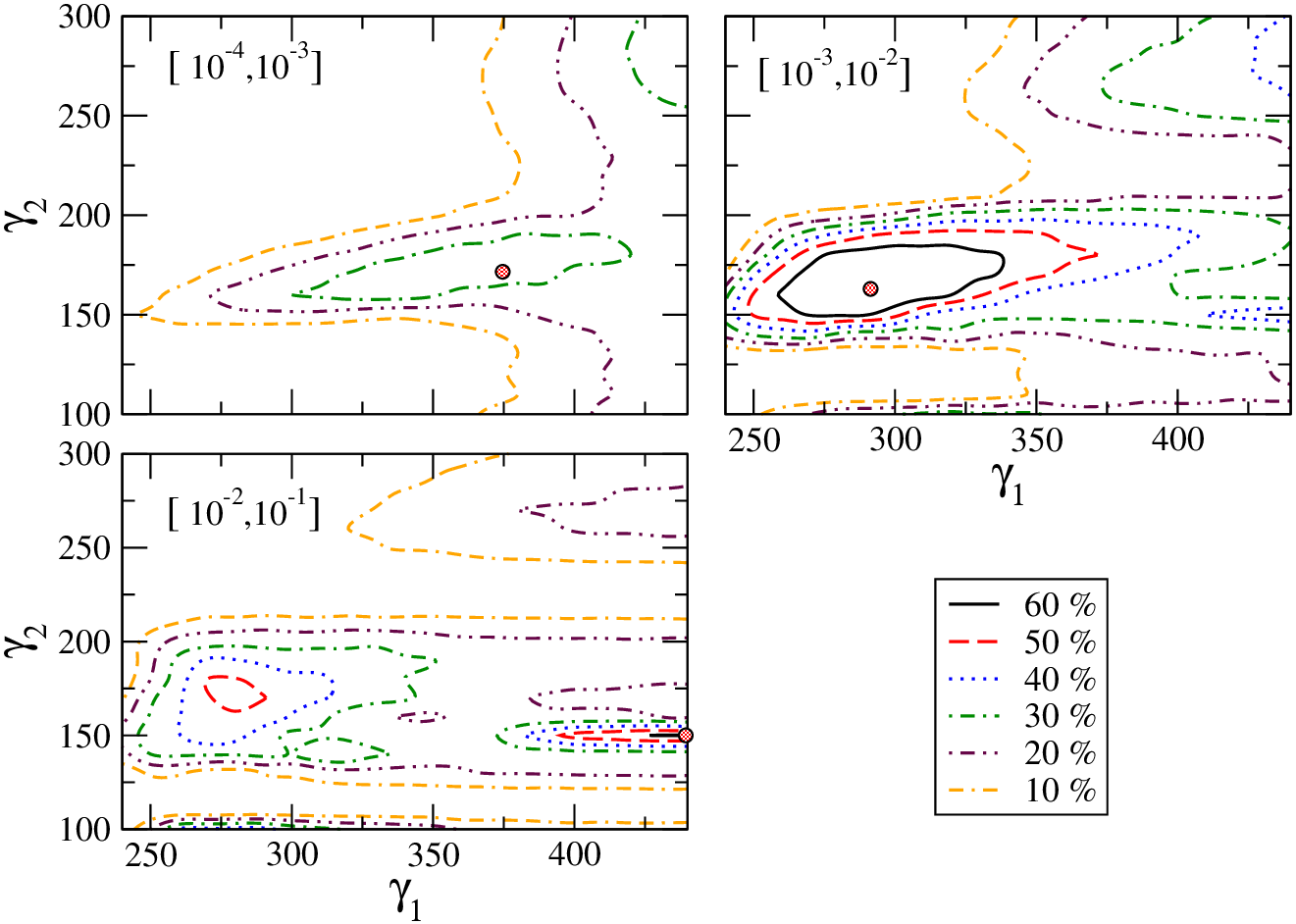}
\end{center}
\caption{For $f=0.3$, the CP-coverage as a function of $\gamma_{1}$ and $\gamma_{2}$ for the individual $\sin^{2}2\theta_{13}$ decades. The red circle indicates the pair that
returns the largest CP-coverage in each analysis.}\label{Fi:decades}
\end{figure}

\begin{figure}
\begin{center}
\includegraphics[width=14cm]{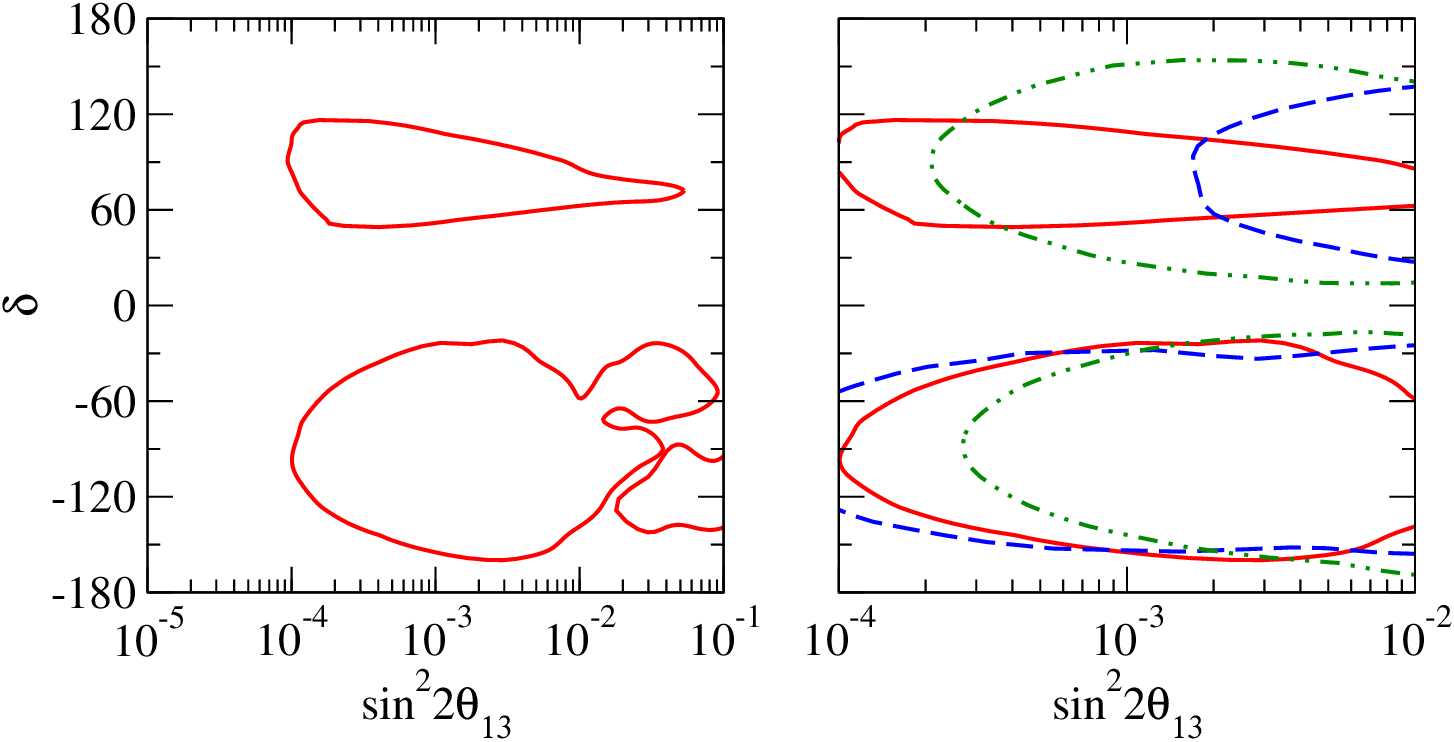}
\end{center}
\caption{CP-violation sensitivity at 99~\% for the pairs \gp $=(370,170)$ (left) and all three boost pairs (right). (370,170) is show in solid red; (440,150) with dashed blue; and
(280,160) as dot-dashed green. }\label{Fi:370_170}
\end{figure}

   The 2-parameter fits for (280,160) are shown in Fig.~\ref{Fi:sensplots_sym} for the same pairs of
true values as before and for normal mass ordering. The good CP-sensitivity for low boost pair is a result of the
CP-features of the appearance probability being out of phase with each other. In particular, the inflexion point of
the sensitivity region for one boost corresponds to the turning point of the other. There are, in general, two such
points; the second corresponds to an energy degeneracy at large \s2. This solution, however, can be ruled out
by near future reactor and accelerator experiments. Note that any energy degeneracy that remains does not cross
the lines \d $=0^{o}$ or \d $= 180^{o}$ and therefore does not interfere with the sensitivity. The relatively poor sensitivity to
CP-violation, is therefore not a consequence of any energy degeneracy; instead it can be attributed to poor \s2
resolution and low event rates.

In Fig.~\ref{Fi:decades}, the analysis is extended to search for further boost pairs of interest through the narrowing of the range of $\sin^{2}2\theta_{13}$ integration. The
range $\sin^{2}2\theta_{13} \in [10^{-4},10^{-1}]$ has been split into three and the scan over the $(\gamma_{1},\gamma_{2}$) plane repeated for each. 
It is apparent that the two pairs 
identified previously are the preferred choices for $\sin^{2}2\theta_{13} \in [10^{-3},10^{-1}]$. For $\sin^{2}2\theta_{13} \in [10^{-4},10^{-3}]$ a new boost pair is identified: 
$(370,170)$ which is explored further in Fig.~\ref{Fi:370_170}. The CP-violation discovery plot (left panel of Fig.~\ref{Fi:370_170}) indicates that discovery down to a 
minimal $\sin^{2}2\theta_{13}=10^{-4}$ can be obtained for both $\delta<0^{\circ}$ and $\delta > 0^{\circ}$. However, this is done at the expense of poor CP-violation discovery for
all $\delta >0^{\circ}$. In the right-hand panel of Fig.~\ref{Fi:370_170}, the CP-discovery analysis is compared to the corresponding analysis for (440,150) and (280,160) for the 
range $10^{-4}< \sin^{2}2\theta_{13}<10^{-2}$, that is to be of interest for a Beta Beam-type facility. The reach is good for $\delta < 0^{\circ}$ with considerable improvement on
the (280,160) pairing. It is clear though that the (370,170) is not a desirable option for $\delta >0^{\circ}$ as two-thirds of those $\delta$, for all $\sin^{2}2\theta_{13}$, cannot be
distinguished from $\delta = 0^{\circ}$ or $\delta=180^{\circ}$. This feature returned a coverage 10~\% smaller from the maximum (280,160) in the first analysis and hence was not 
identified there.

Hereafter, the analysis will concentrate on the two pairs (280,160) and (440,150). The first is of interest if one requires to the electron capture machine to be configured 
to achieve a broad reach down to $\sin^{2}2\theta_{13}\sim$ few $\times 10^{-4}$, whilst the second performs better close to the current limit on $\sin^{2}2\theta_{13}$ and the 
range to be explored by near-future experiments. In the next section, these two facilities will be examined further through the variation of the number of useful decays and the 
level of atmospheric background.

\section{Useful decays and atmospheric backgrounds}\label{S:bck}

   In the previous section, the scan of the $(\gamma_{1},\gamma_{2}$) plane was carried out assuming that $10^{18}$ useful
ion decays per year will be available and that the number of atmospheric background events is zero. As discussed
in Sec.~\ref{S:concept}, this useful decay rate is challenging and would require technology and R\&D beyond that required for the
standard Beta Beam ions. Zero atmospheric backgrounds is not a realistic assumption; with no energy reconstruction
nor background rejection, every atmospheric event that passes the cuts will be misidentified as an $\nu_{\mu}$ appearance
event. Assuming a duty factor of $10^{-3}$, i.e. only 0.1~\% of the decay ring is filled with ions, there will about 0.03
atmospheric neutrino events per kton-year~\cite{BB_SPS}. With a 440 kton detector this amounts to 13.2 events per year. This
is a constant that will need to be added to the number of events for all count rates in the simulation. The effect of
the atmospheric background is felt at the sensitivity boundary. In these regions, the number of events is the same
order as the atmospheric background. The inclusion of the atmospheric background reduces the value of the $\chi^{2}$ and
pushes the boundary inwards. This can be seen trivially with the Gaussian form of a $\chi^{2}$ . For true event rate $n_{i}$, test
event rate $\zeta_{i}$ and overall systematic $f_{\rm sys}$, the Gaussian $\chi^{2}$ is given by
\begin{equation}
\chi^{2}= \sum_{i}\frac{(n_{i}-\zeta_{i})^{2}}{\zeta_{i}+(f_{\rm sys}\cdot\zeta_{i})^{2}}~.
\end{equation}
Inclusion of a constant background, $B$, is the replacement $n_{i} \rightarrow n_{i} + B$ and similarly for $\zeta_{i}$. This constant
translation of the event rates leaves the numerator of the $\chi^{2}$ unchanged whilst increasing the denomenator. The
result is that the $\chi^{2}$ is reduced.

   The physics reach of the electron capture beam cannot be stated until the atmospheric background is included.
Further, it is not known what is a realistic useful decay rate per year - it is reasonable to assume that $10^{18}$ may well
be close to an upper limit though. A technique to improve the decay rate is to increase the duty factor which in turn
introduces a larger atmospheric background rate. Since the useful decay rate is not known, the approach requested
by the experimentalists~\cite{mats} is to treat the useful decay rate and number of atmospheric events as independent.

This analysis has been carried out using the CP-coverage measure and $\sin^{2}2\theta_{13}\in [10^{-5},10^{-1}]$, as previous. Note, however, that its use here is for a different purpose. Whereas before the goal was to identify boost pairs of interest, here it is used to monitor the area change of the 99~\% confidence region. As the 
decay rate drops and/or the atmospheric neutrino level is increased, this region will shrink by shifting inwards. The CP-coverage allows one to adequately measure this change.

\begin{figure}
\begin{center}
\includegraphics[width=7cm]{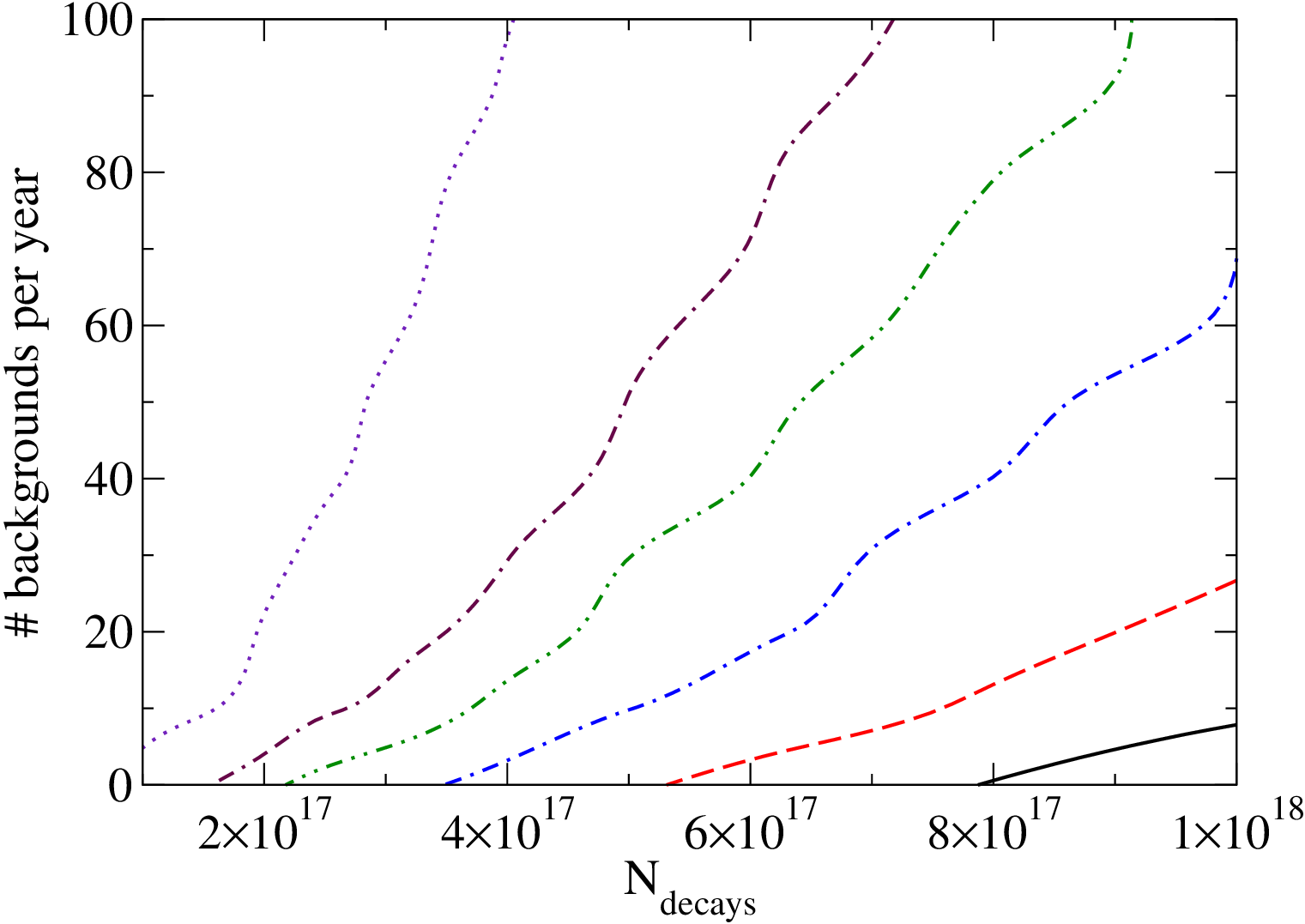}
\includegraphics[width=7cm]{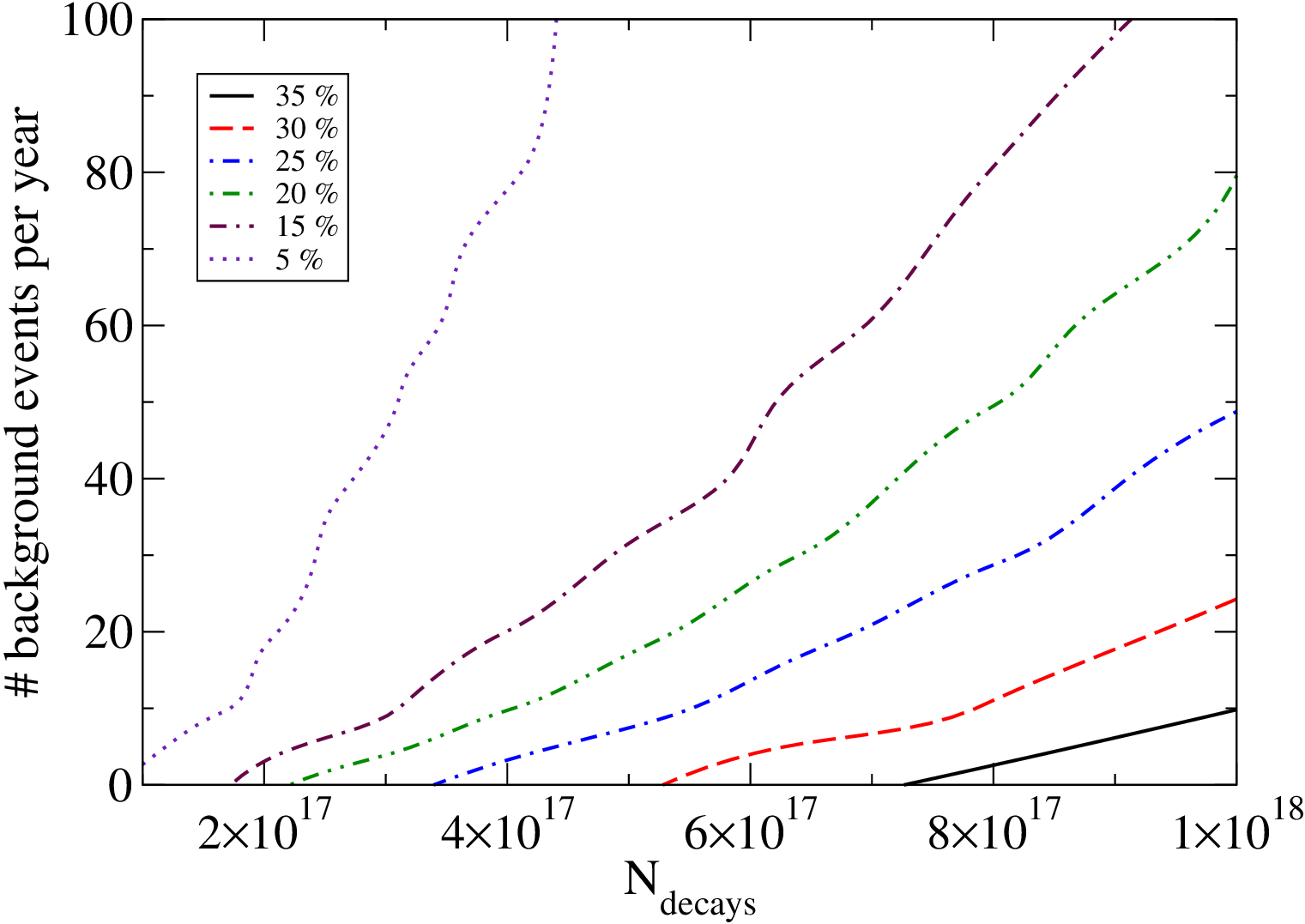}
\end{center}
\caption{CP-violation coverage for $f=0.3$ for \gp $=(440,150)$ (left) and \gp $=(440,150)$ (right) as a function of the number of useful decays and number of atmospheric backgrounds per year.}\label{Fi:dec_atm}
\end{figure}

The results are presented in Fig.~\ref{Fi:dec_atm}. The two pairs (440,150) and (280,160)
with $f = 0.3$ have been re-simulated incorporating the number of background events per year and varying the number
of useful ion decays. It is seen that the electron capture machine, like all Beta Beam type machines, is statistics
dominated; the coverage of the (\s2,~\d) plane is reduced considerably with drops in the decay rate. For example,
if the average number of atmospheric background events per year is 10, reducing the useful decay rate by a factor of
2 reduces the coverage by almost 10~\% of the plane for both cases. When the coverage is only 35~\% for $10^{18}$ useful
decays, this is a substantial drop. On the other hand, if one instead opts to regain that factor of 2 by loosening the
duty factor, i.e. double the number of atmospheric neutrino events, the reduction is only 2-3~\%. This observation seems to agree with a previous study for a CERN-Frejus
$^{18}$Ne and $^{6}$Be Beta Beam~\cite{mauro_atm} where it was demonstrated that a loosening of the duty factor does not lead to the substantial reductions in physics return as
one might assume. 

\begin{figure}
\begin{center}
\includegraphics[width=7cm]{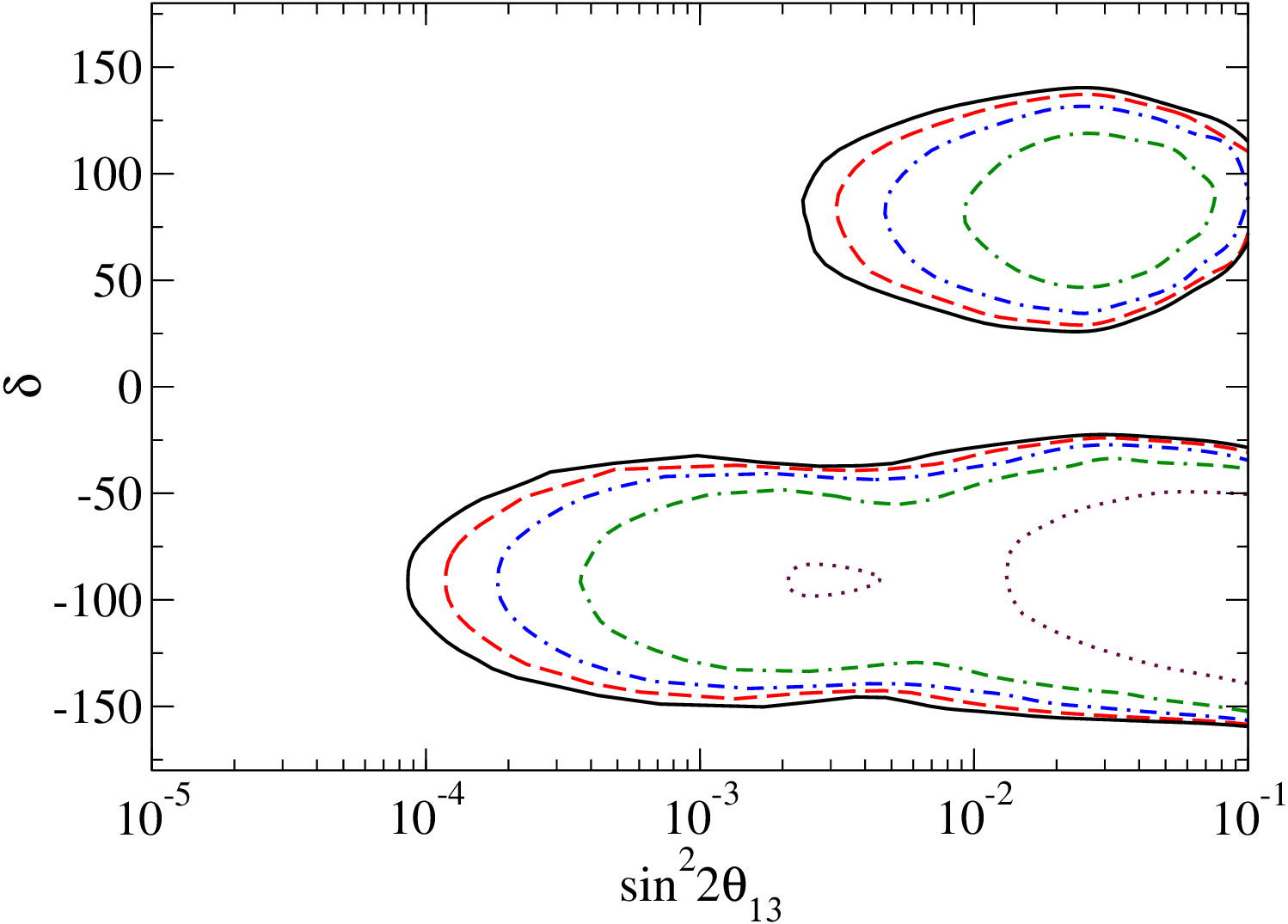}
\includegraphics[width=7cm]{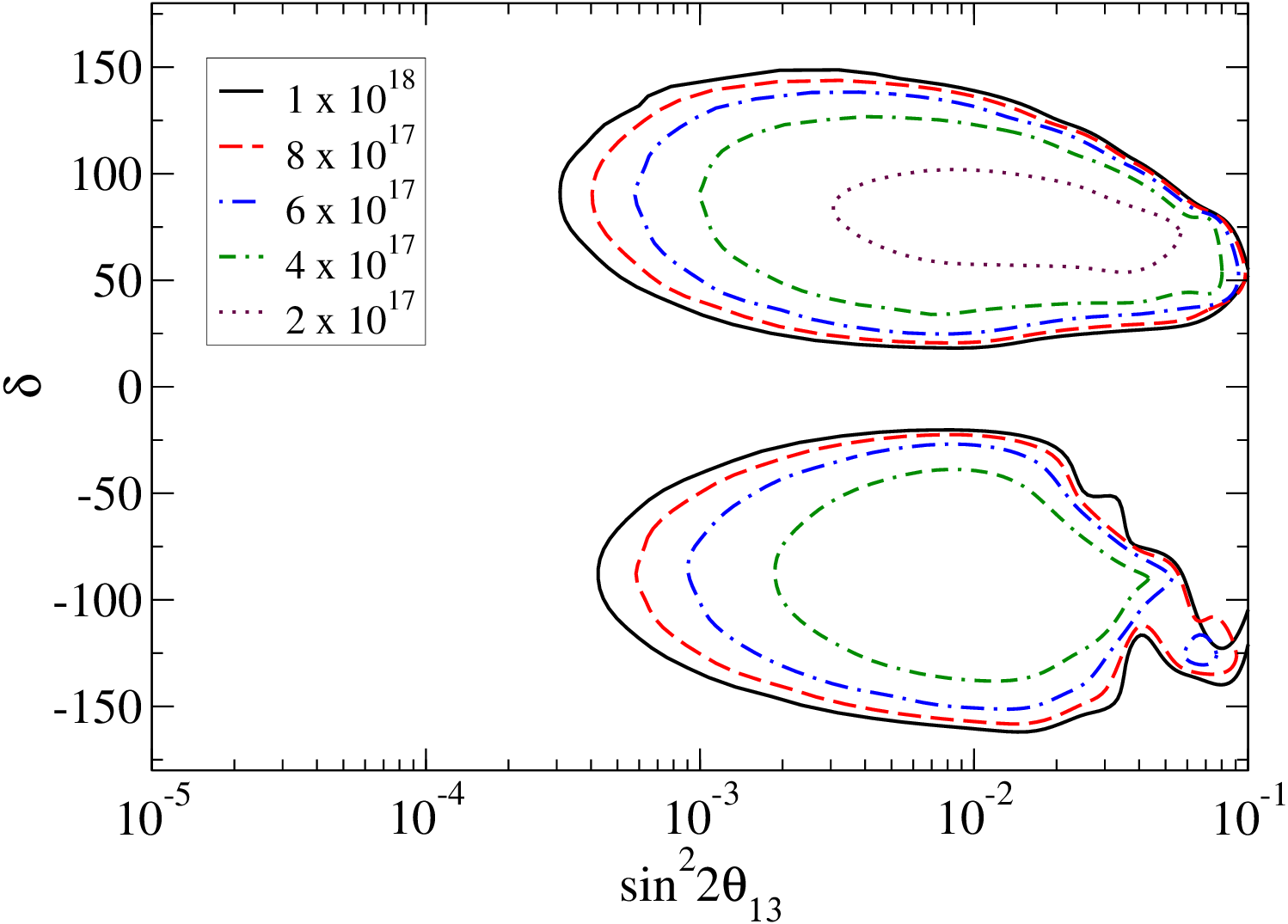}
\end{center}
\caption{CP-violation sensitivity at 99~\% confidence level for $f=0.3$ and $N_{\rm atm}=13.2$ per year for \gp =(440,150) (left) and (280,160) (right). In both cases, the sensitivity contour has been constructed for a range of useful decay rates.}\label{Fi:decayrates}
\end{figure}

\begin{figure}
\begin{center}
\includegraphics[width=7cm]{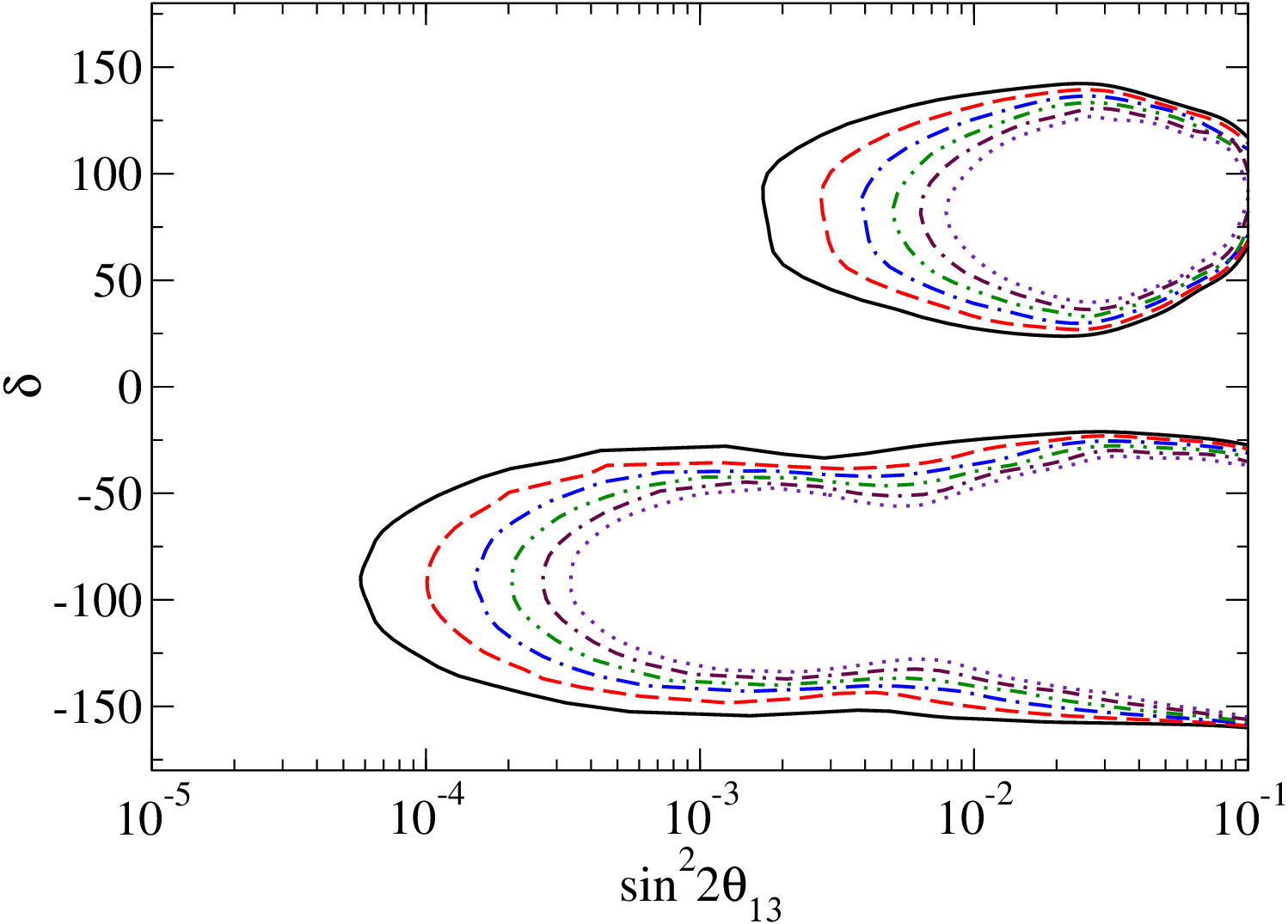}
\includegraphics[width=7cm]{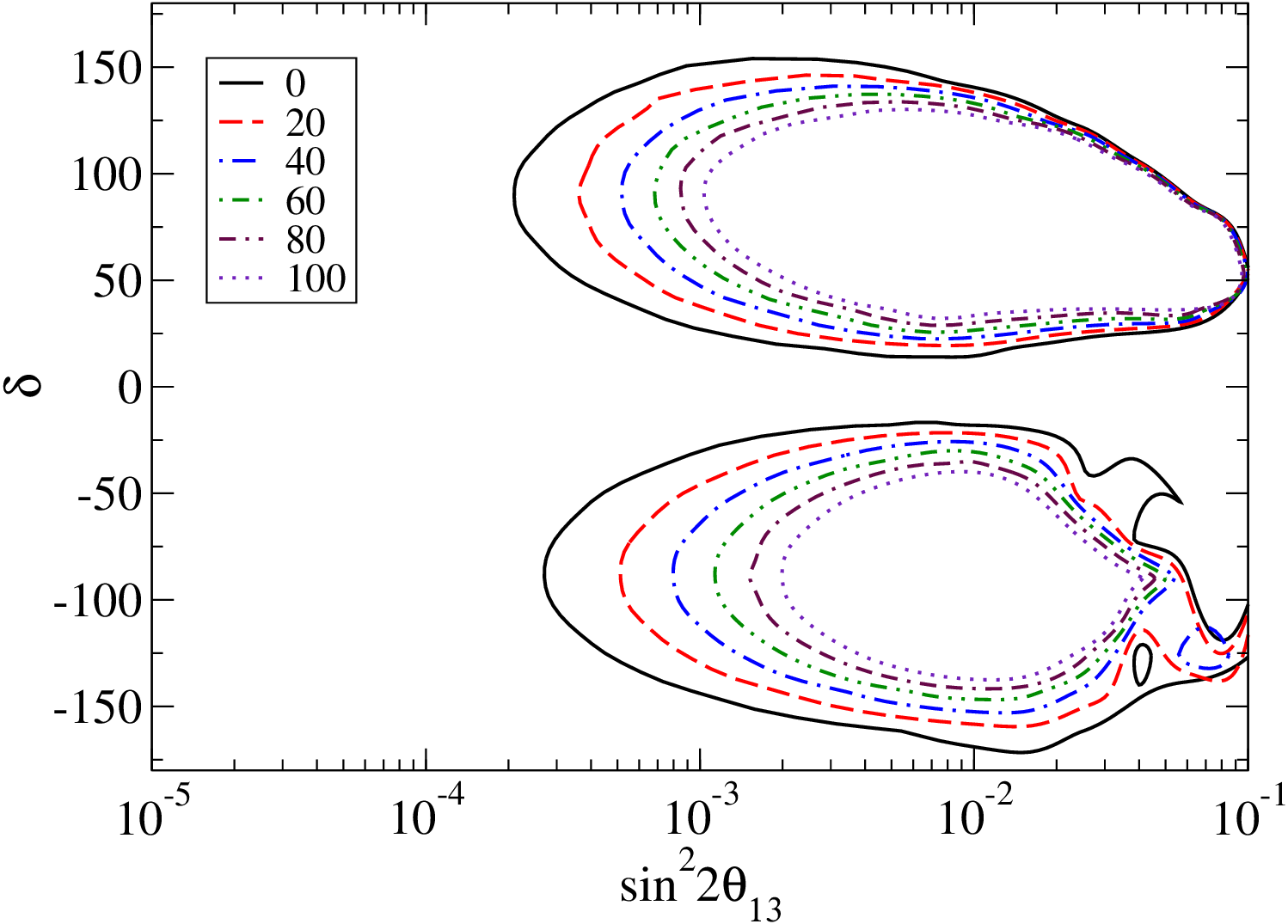}
\end{center}
\caption{CP-violation sensitivity at 99~\% confidence level for $f=0.3$ and $N_{\rm ions}=10^{18}$ per year for \gp =(440,150) (left) and (280,160) (right). In both cases, the sensitivity contour has been constructed 
for a range of atmospheric neutrino events.}\label{Fi:atmbck}
\end{figure}

   To see the results more explicitly, in Fig.~\ref{Fi:decayrates} CP-sensitivity plots have been constructed (for both cases) including
a constant atmospheric background of 13.2 events per year for a range of useful ion decay rates. The choice of boost
pairs presented here provide excellent resolution of the energy degeneracy. As a consequence, a lower event rate merely
reduces to overall sensitivity of the setup, rather than let degeneracies ruin the sensitivity. The first manifestation
of any degeneracy is for $N_{\rm ions} = 2 \times 10^{17}$ per year for (440,150) at \s2 $\sim 10^{-2}$ and \d  $< 0^{o}$ . In Fig.~\ref{Fi:atmbck} this exercise has
been repeated but with $N_{\rm ions} = 10^{18}$ per year kept fixed and the atmospheric event rate varied between 0 and 100 events
per year. Here the reduction in sensitivity is less extreme with no degeneracy appearing. The effect of the background is only felt through the denomenator of the $\chi^{2}$ so the facilities ability to resolve
degeneracy is largely unaffected. An increased background therefore pushes the sensitivity boundary at a given confidence level inwards towards the set of pairs \td~where the event rate and background rate are
of the same order.    

These plots indicate that the physics reach of the electron capture machine is a struggle between sourcing a plentiful
useful decay rate whilst keeping the atmospheric background to a minimum. With binning of the neutrino signal,
the constant atmospheric neutrino rate can be reduced whilst keeping the useful decay rate fixed. There are two
drawbacks to this possibility:
\begin{enumerate}
\item The use of QE-events would be necessary
\item An event reconstruction efficiency needs to be included
\end{enumerate}
The electron capture machines need to use large Water Cerenkov detectors to achieve a competitive event rate: the
cross-sections are small at the lower energies studied at short baselines. The energy reconstruction of neutrino events
at a Water \v{C}erenkov is through selection of quasi-elastic events with an efficiency of about 60-70~\%. This
is in addition to the loss of events from not using all the charge-current events.
There will also be a small further loss owing to finite energy resolution. Binning the signal will
reduce background but the reduction in events is essentially equivalent to using a smaller useful decay rate. There is
therefore likely to be no gain in using this approach.

Alternative strategies to reduce the atmospheric background rate would be to either introduce cuts on the signal or to use directional information. Since the
neutrinos are at a known energy, one can impose cuts on the signal for larger and smaller energies. The atmospheric events are skewed towards low energies, so this
would work best for large ion boosts. At high energies, the neutrino direction is correlated with the observed lepton direction. This gives a means to separate
the beam signal from the approximately isotropic atmospheric signal. However, any gain from these extra reductions will probably be offset by an increase in the
duty factor in an attempt to achieve a useful decay rate.

\section{Mono-energetic anti-neutrinos}\label{S:BBD}

   The electron capture machine setups with the largest CP-violation coverage are either asymmetrical in $\delta$ or symmetrical, 
but with a larger \s2 limit. It is clear that unless vast improvements in the available number of useful decays
per year are achieved or envisaged, the physics reach of electron capture machines cannot compete with the standard
Beta Beams over equivalent baselines which can rule out CP-conservation down to \s2 $\sim 10^{-4}$ and are close to
symmetric in \d. The reason for this is the use of a neutrino run only. The need for information complementary to
first oscillation maximum data, forcing runs at smaller boosts, has reduced the overall event rate across the whole experimental
run. One option to explore is whether it is possible to construct a facility with a mono-energetic anti-neutrino flux
at first oscillation in combination with an equivalent neutrino flux from electron capture. The two fluxes would be
complementary with the need for additional runs at lower energies unnecessary. The event rate across the entire run
of any experiment would be higher, with less susceptibility to variations in the useful decay rate. One process put
forward as a complementary source of mono-energetic anti-neutrinos is the bound beta decay process (BBD)~\cite{CPeven}.

   In a continuum beta decay (CBD) the electron is virtually always ejected from the atom with capture in the outer
orbitals strongly suppressed due to weak bindings and small wavefunction overlaps. However, for a fully ionised atom,
electron capture in the K-shell (the bound beta decay process (BBD)) can have a substantial branching ratio, provided the transition Q-value is sufficiently
small and proton number of the ion is large:
\begin{equation}
\frac{\Gamma_{B}}{\Gamma_{\beta}} = \frac{Q_{B}^{2}\vert \psi_{n}(0)\vert^{2}}{f(Q_{C},Z)}~,
\end{equation}
where
\begin{equation}
f(Q_{C},Z)=\int_{m_{e}}^{Q_{C}+m_{e}}E\sqrt{E^{2}-m_{e}}(Q_{C}-E)^{2}F(Z,E)dE
\end{equation}
is the integral over phase space for CBD and $\psi_{n}(0)$ is the wavefunction of the electron in the nth orbital. For a fully ionised atom, the Q-value for the
BBD process is given by
\begin{alignat}{1}
Q_{B}^{Z+}&=Q_{C}\:\:+\vert B_{n}(I')\vert - \vert \Delta B^{\rm tot}(I',I)\vert~, \notag \\
         &= Q_{C}^{Z+}+\vert B_{n}(I')\vert~,
\end{alignat}
where $\vert \Delta B^{\rm tot}(I',I)\vert$ is the difference in binding energies of the complete parent and daughter atoms, and $\vert B_{n}(I')\vert$ is the
binding energy of the electron captured into the orbital $n$ of the daughter nucleus. Specifically, the Q-values for the BBD and CBD channels are separated by 
$\vert B_{n}(I)\vert$ in the rest frame of the nucleus.

In the last couple of years, use of BBD has been forwarded for very short baseline experiments~\cite{Minakata} searching for oscillations of so-called `Mossbauer 
neutrinos'~\cite{Moss}. The possible 
reaction rates for a range of candidate ions sourced from a low-boost Beta Beam has also be discussed in~\cite{Oldeman}. 
Here the possibility of using BBD as a neutrino source for a long baseline neutrino oscillation experiment is considered as was suggested in~\cite{CPeven} where the authors proposed the use of 
ions that can BBD as well as electron capture and CBD. The paper defined the `CP-evenness' as
\begin{equation}
\eta(E,\gamma) = \frac{\mathcal{F}(\nu_{e};E)\sigma(\nu_{\mu};E)-\mathcal{F}(\bar{\nu}_{e};E)\sigma(\bar{\nu}_{\mu};E)}
                      {\mathcal{F}(\nu_{e};E)\sigma(\nu_{\mu};E)+\mathcal{F}(\bar{\nu}_{e};E)\sigma(\bar{\nu}_{\mu};E)} ~,
\end{equation}
where $\mathcal{F}$ are the fluxes and $\sigma$ the cross-sections at definite energies. Since $\mathcal{F}(\nu)\sigma(\nu)$ is the un-oscillated number of events
at the detector, it was suggested that an optimum neutrino beam is one with $\eta\approx0$ (equal un-oscillated neutrino ans anti-neutrino events). The feasibility
of this approach and the combination of an electron capture machine and bound beta decay machine was discussed in~\cite{BBDme}. 
Below I will summarise the main reasons
why a machine based on the BBD process is not a realistic proposition.

\begin{enumerate}
\item \emph{The branching ratios are too small}. For a fully ionised ion to have a BBD substantial branching ratio requires
a large proton number and a very small Q-value. There is always a substantial contribution from the CBD. A
scan of a database searching for suitable ions with half-lives in the range $0.5\:\: {\rm sec} < t_{1/2} < 8\:\: {\rm min}$ and a dominant
decay channel returned very little. The largest BBD branching ratio found is for $^{207}$Tl$^{81+}$: 12~\%. Therefore to
source an equivalent anti-neutrino event rate requires an extra factor of 10 useful decays, compared to electron
capture decays, and a further factor of 3 to compensate for the lower anti-neutrino cross-section. The ions
suggested for electron capture and BBD in~\cite{CPeven} had much lower branching ratios around 1~\%.
\item \emph{Magnetised detectors are necessary for ions with both electron capture and BBD}.
For ions that
electron capture and BBD, it is necessary to separate $\mu^{-}$ and $\mu^{+}$ events at the detector. For MIND detectors~\cite{ISS},
the threshold is too high for the intermediate baselines of Europe. Magnetised Liquid Argon or Totally Active
Scintillator detectors would be necessary. However, the previous comment indicates that a factor of 100 is missing
to get a useful number of mono-energetic neutrinos of order $10^{18}$, so this is a mute point.
\item \emph{Very large boosts are required to separate BBD and CBD channels in the laboratory frame.}
CBD will always be the dominant channel. It is therefore necessary to separate this out from the BBD channel
otherwise the mono-energetic nature of the BBD is not being exploited. Failure to do this would result in a
Beta Beam using a high proton number ion with only the endpoint of the flux in use. The energy split between CBD and BBD in
the laboratory frame is $E = 2\gamma B_{1}(Y)$, for boost $\gamma$ and decay to the K-shell of daughter ion Y. For example, $^{207}$Tl$^{81+}$ has
$B_{1}(Y)$ = 99 keV. For a boost of $\gamma = 400$, this corresponds to a required binning and energy resolution $\Delta E < 80$
MeV, which is tough. More conservative binnings of 150-200 MeV and greater are often considered. Boosts in
excess of 800 will be required to have an empty bin between the CBD and BBD fluxes in these cases. Such
fluxes require the LHC and will diminish the useful decay rate further due to its leading order $1/\gamma$ dependence. For all ions with smaller proton numbers,
the experimental task will be harder owing to the smaller binding energies.
\end{enumerate}
In short, the implausibility of achieving $\mathcal{O}(10^{18})$ useful mono-energetic anti-neutrinos per year and separating out the BBD
anti-neutrinos from the CBD anti-neutrinos does not make a Beta Beam-type facility based on BBD a viable future
facility. This does not mean that anti-neutrinos cannot accompany an electron capture beam. That can be achieved
by combining an electron capture machine with the standard anti-neutrino ions; $^{6}$He and $^{8}$Li. This is left for a different
study.

\section{Summary and conclusions}\label{S:con}

   A electron capture beam provides a means to source long baseline neutrino oscillation experiments with a mono-energetic 
$\nu_{e}$ flux. By altering the Lorentz boost of the ion, one can freely choose the laboratory neutrino energy up
to some maximum determined by the Q-value and the maximum energy of the accelerator. Since a single neutrino
energy is insufficient to determine the unknown neutrino mixing parameters and resolve any degeneracy, it is necessary
to include at least two boosts in any physics strategy. In the present article, a previous work~\cite{catalina2} examining the physics
reach of an electron capture beam sourced from CERN and directed towards a 440 kton fiducial mass Water Cerenkov
detector was extended to include the study of the choice of boost, the relative run times of each boost, the number
of useful ion decays, and the number of atmospheric neutrino events. The use
of neutrinos only, but with multiple energies introduces a variant on the intrinsic degeneracy that is not present in
studies that use both neutrino and anti-neutrinos. This `energy degeneracy' was discussed and
shown to present in electron capture beam facilities unless the two oscillation phases $\Delta m_{31}^{2}/4E$ are separated by $\pi$
and the CP-phase \d~takes a specific value.

   In the first instance, the number of useful decays was fixed to $10^{18}$ per year, and atmospheric backgrounds were
neglected.  Little variation was found for the relative run times; however, two
regions of the \gp~ plane returned large CP-coverage. The naive pairing of (almost) first and second oscillation
maximum had a CP-coverage of approximately 35~\% but was asymmetric in $\delta\rightarrow -\delta$. For $\delta < 0$, this choice could
rule out CP-conservation down to \s2 $= 10^{-4}$, but only \s2$\sim 2\cdot 10^{-3}$ for $\delta > 0$. There exists a slightly
larger coverage for the pair (280, 160) with run time fraction $f = 0.3$. This choice returns a (roughly) symmetrical
CP-sensitivity region with CP-conservation ruled out down to \s2$\sim 4\cdot 10^{-4}$.

   No R\&D has been carried out on the feasibility of sourcing $10^{18}$ useful decays per year for high proton number ions.
A brief study adapting Beta Beam codes indicated that the decays are too slow and large tune-shifts would need to
be accommodated. With existing technology, the number of useful decays is two orders of magnitude too small. At
best, $10^{18}$ useful decays per year appears to be overly optimistic and should be considered as the hard limit on the
yearly rate. Some of this problem could be alleviated by keeping large numbers of electrons bound to the nucleus
whilst still having a large enough charge-to-mass ratio to reach the sought boost.

   In a standard Beta Beam, a restrictive duty factor of $S_{f}\sim 10^{-3}$ is normally taken as necessary to reduce the
atmospheric neutrino flux to a level that does not destroy the sensitivity. The next step, therefore, was to examine
the CP-sensitivity as a function of both the yearly useful decay rate and the number of atmospheric neutrino events
in the appearance sample. In both cases, the results were more volatile to small changes in the useful decay rate than
the atmospheric background. For instance, taking $N_{\rm ions} = 10^{18}$ and $B_{\rm atm} = 10$ events per year, the CP-coverage is approximately
35~\% in both cases. If the decay rate is dropped by a factor of 2, then the coverage falls to about 23~\%. If instead
the atmospheric background is doubled, then the coverage only drops to about 32~\%. The principle reason for this
difference is the total event has manifestations in both the numerator and denomenator of the $\chi^{2}$ function whereas
the background only affects the denomenator. The discrepancy between two \td~is felt though the difference
squared in the numerator. Drops in the total event rate can therefore lead to degenerate solutions dropping below
the required statistical significance of the test thus reducing the sensitivity in a manner beyond simply a scaling of
the $\chi^{2}$. The background merely reduces the $\chi^{2}$ without interfering with its ability to rule out degenerate solutions.
Its manifestation is simply to push the sensitivity contours inwards.

\begin{figure}
\begin{center}
\includegraphics[width=7cm]{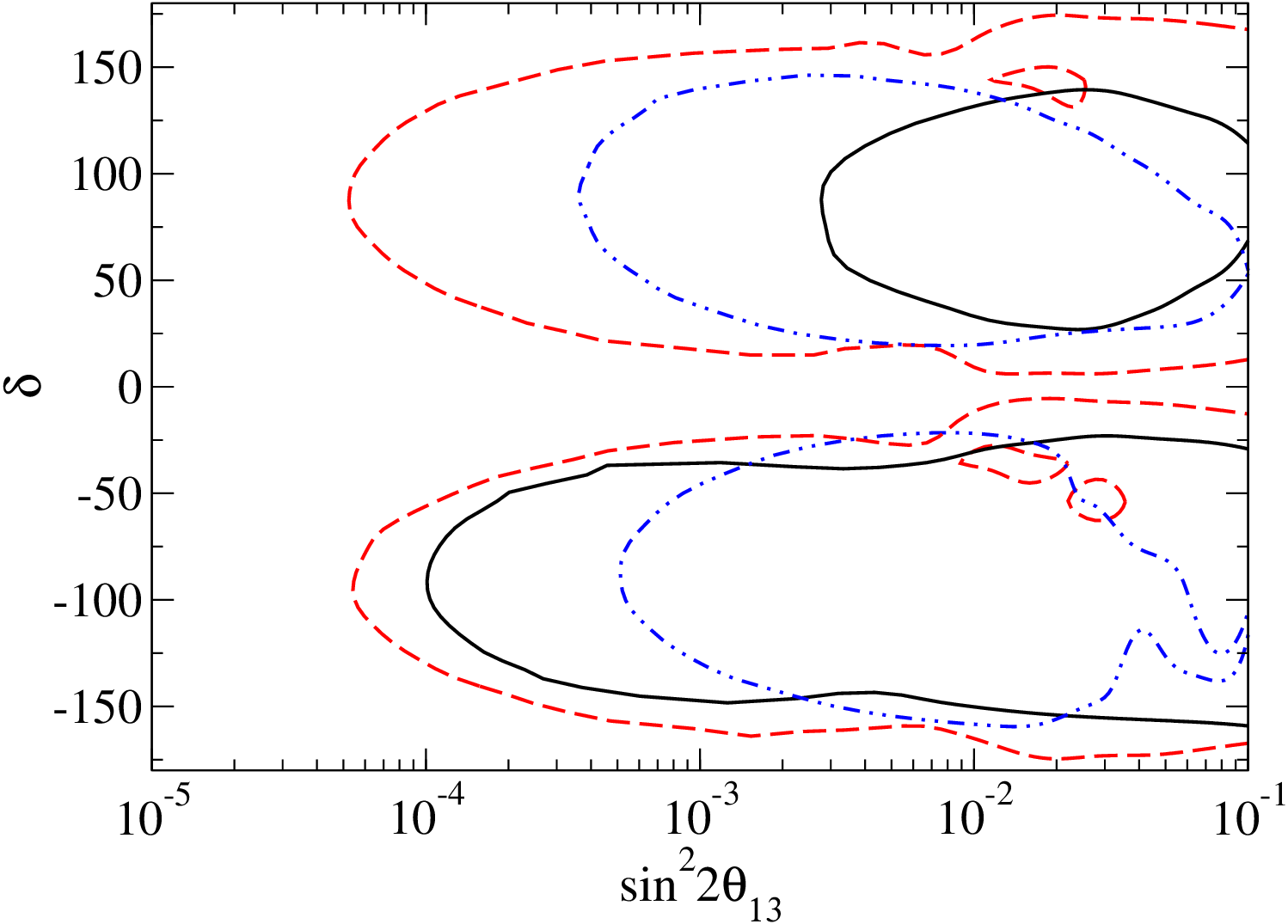}
\end{center}
\caption{CP-violation sensitivity at 99~\% confidence level for $f=0.3$ and $N_{\rm atm}=20$ per year for \gp =(440,150) (black solid) and (280,160) (blue dot-dashed). The red
dashed line displays the physics return for a standard Beta Beam directed along the CERN-Canfranc baseline with a boost $\gamma=350$ for both neutrinos and anti-neutrinos.}\label{Fi:comparison}
\end{figure}
 
   Ultimately, the limiting factor for an electron capture machine is the number of useful decays. The advantages of
using a mono-energetic neutrino beam is the precise knowledge of the neutrino energy and ability to choose which
part of the oscillatory structure of the appearance probability is to be explored. Ironically these are also the main
disadvantages. Since a single energy is insufficient to extract the unknown mixing parameters, it is necessary to
run for substantial periods of time at low energies where the event rate is small. The availability of an equivalent
anti-neutrino beam would remove this necessity. Ions that bound beta decay could source such a beam; however,
this will not be practical with the available technology.

   The feasibility of an electron capture beam is an open question. However, the most optimistic parameterisation
considered in this article is likely to be beyond the hard limits imposed by available technology. It must be conceded
that if \t13 were to be very small, the electron capture will not be a competitive facility, unless a fast decaying ion that
can be produced in large quantities is found. This can be seen explicitly in Fig.~\ref{Fi:comparison} 
where the CP-discovery for the boost pairs (280,160) and (440,150), and 20 atmospheric events 
per year, are compared to the $\gamma=$ 350,350 Beta Beam introduced in~\cite{BB_SPS} with the baseline shortened to CERN-Canfranc, as in this article. The simulation of 
this facility uses $1.1\times 10^{18}$ neutrinos per year sourced from the decay of $^{18}$Ne, and $2.9\times 10^{18}$ anti-neutrinos per year sourced from decays of $^{6}$He. 
The run time is 10 years with an equal split between neutrino and anti-neutrino running. We see that for a (440,150) electron capture facility, the physics return is poorer but 
relatively competitive for $\delta < 0^{o}$. For these $\delta$, the (280,160) facility is only competitive for $\sin^{2}2\theta_{13}\in [10^{-3},10^{-2}]$. For $\delta > 0^{o}$, the 
(440,150) facility is outperformed by the standard Beta Beam by 2 orders of magnitude whilst, again the (280,160) facility is only competitive for 
$\sin^{2}2\theta_{13}\in [10^{-3},10^{-2}]$.

In conclusion, an electron capture machine is unlikely to be a competitive facility for the measurement of CP-violation. Taking into account technological challenges and 
restrictions, the CP-violation reach of a possible electron capture machine is probably
going to be limited to \s2 $> 10^{-3}$. This is not to say that an electron capture machine is a redundant option. The sensitivity suffered from the need to run
for substantial periods of time at low energies. If the electron capture machine is used in combination with another facility, this will be become
unnecessary. In such a scenario the role of the machine would be to complement a facility that has both neutrino and anti-neutrino fluxes and/or coverage
of a large energy range. In this case, the electron capture machine could be run for 100~\% of the time at a well chosen energy to compensate for any
deficiency in the physics reach of the other facility.

\acknowledgments

The author would like to thank Silvia Pascoli, Sergio Palomares-Ruiz and Jose Bernabeu for discussions and suggesting improvements to the manuscript; 
Mats Lindroos for advice on the production and acceleration; Dan Roythorne for a discussion on the simulation strategy; and the CERN ISOLDE group for their
hospitally on a number of occasions early in the study.
  This work was carried out under a STFC studentship and additionally supported by 
 the European Community under the European Commission Framework Programme 7 Design
studies: EUROnu, Project Number 212372; and LAGUNA, Project Number 212343. 
The EC is not liable for any use that may be made of the information
contained herein.

\end{document}